\documentclass[aps,prd,showpacs,superscriptaddress,onecolumn,raggedfooter,raggedbottom]{revtex4}   

\usepackage{amssymb}
\usepackage{amsmath}
\usepackage{amsthm}
\usepackage{graphicx}
\usepackage{subfigure}
\usepackage{color}
\usepackage{mathrsfs} 
\usepackage{paralist}

\usepackage{hyperref}
\hypersetup{hidelinks}
\usepackage{soul}

\usepackage[capitalize,nameinlink]{cleveref}
\crefname{section}{section}{sections}
\crefname{subsection}{subsection}{subsections}
\Crefname{section}{Section}{Sections}
\Crefname{subsection}{Subsection}{Subsections}
\Crefname{figure}{Figure}{Figures}
\crefformat{equation}{\textup{#2(#1)#3}}
\crefrangeformat{equation}{\textup{#3(#1)#4--#5(#2)#6}}
\crefmultiformat{equation}{\textup{#2(#1)#3}}{ and \textup{#2(#1)#3}}
{, \textup{#2(#1)#3}}{, and \textup{#2(#1)#3}}
\crefrangemultiformat{equation}{\textup{#3(#1)#4--#5(#2)#6}}%
{ and \textup{#3(#1)#4--#5(#2)#6}}{, \textup{#3(#1)#4--#5(#2)#6}}{, and \textup{#3(#1)#4--#5(#2)#6}}
\Crefformat{equation}{#2Equation~\textup{(#1)}#3}
\Crefrangeformat{equation}{Equations~\textup{#3(#1)#4--#5(#2)#6}}
\Crefmultiformat{equation}{Equations~\textup{#2(#1)#3}}{ and \textup{#2(#1)#3}}
{, \textup{#2(#1)#3}}{, and \textup{#2(#1)#3}}
\Crefrangemultiformat{equation}{Equations~\textup{#3(#1)#4--#5(#2)#6}}%
{ and \textup{#3(#1)#4--#5(#2)#6}}{, \textup{#3(#1)#4--#5(#2)#6}}{, and \textup{#3(#1)#4--#5(#2)#6}}
\crefdefaultlabelformat{#2\textup{#1}#3}

\begin{document}

\title{Fractional Solitons: A Homotopic Continuation from the Biharmonic to the Harmonic $\phi^4$ Model}

\author{Robert J.\ Decker}
\affiliation{Mathematics Department, University of Hartford, 200 Bloomfield Ave., West Hartford, CT 06117, USA}

\author{A.\ Demirkaya}
\affiliation{Mathematics Department, University of Hartford, 200 Bloomfield Ave., West Hartford, CT 06117, USA}

\author{T.J. Alexander}
\affiliation{Institute of Photonics and Optical Science (IPOS), School of Physics, The University of Sydney, NSW 2006, Australia}

\author{G.~A.\ Tsolias}
\affiliation{Department of Mathematics and Statistics, University of Massachusetts,Amherst, MA 01003-4515, USA}

\author{P.~G.\ Kevrekidis}
\affiliation{Department of Mathematics and Statistics, University of Massachusetts,Amherst, MA 01003-4515, USA}

\begin{abstract}
In the present work we explore the path from a harmonic to a biharmonic PDE of Klein-Gordon type from a continuation/bifurcation perspective.
More specifically, we make use of 
the Riesz fractional derivative as a 
tool that allows us to interpolate between these
two limits.  We illustrate, in particular,
how the coherent kink structures existing in these models transition from
the exponential tail of the harmonic operator
case, via the power-law tails of intermediate
fractional orders, to the oscillatory exponential
tails of the biharmonic model. Importantly,
we do not limit our considerations to the
single kink case, but extend to the kink-antikink
pair, finding an intriguing cascade of saddle-center 
bifurcations 
happening  exponentially close to the 
biharmonic limit. Our analysis clearly explains
the transition between the infinitely many 
stationary soliton pairs of the biharmonic case
and the absence of even a single such pair in the
harmonic limit. The stability of the different
configurations obtained and the associated dynamics
and phase portraits are also analyzed.
\end{abstract}

\maketitle


\section{Introduction}

The past decade has seen tremendous growth
in interest in fractional
calculus models, due to the
number of potential applications of these models. 
Associated proposals range from fractional diffusion in biological systems~\cite{IonescuCNSNS2017}, to epidemiological modelling of the spread of measles~\cite {qureshi2020real}, to the spread of computer viruses~\cite{singh2018fractional, azam2020numerical}. Other fields
of interest include optical sciences~\cite{malomed},
nonlinear waves~\cite{cuevas} and
economics~\cite{ming2019application}. This volume of studies and
associated applications is now reflected
in a number of reviews and books; see, 
e.g.,~\cite{podlubny1999fractional,Samko}.

Importantly, the field also features developments in the fundamental mathematical analysis of the 
relevant operators~\cite{yavuz2020comparing, saad2018new}, however this area is nowhere near
fully explored~\cite {ortigueira2021two, ortigueira2021bilateral}. 
 While there are numerous
 proposed definitions of fractional derivatives
 with notable similarities, but also differences,
arguably the Riesz fractional derivatives
provide a quite  promising theoretical framework for approaching
physically relevant
settings~\cite{muslih2010riesz}.
For instance, such derivatives have been
identified as suitable continuum limits
for systems of particles involving
long-range interactions~\cite{tarasov2006fractional}.
 
Despite the promise of applications, there remains a dearth of indisputable experimental realizations of a system consistent with a fractional calculus theoretical framework.  A direction of
considerable interest that may yet allow for a link between theory and experiment is in nonlinear optics where marked success has been achieved in controlling the dispersion order~\cite{BlancoRedondoNC2016,RungeNP2020}.  
It should be noted that the fractional Schr{\"o}dinger equation
was proposed in optics in the work of~\cite{Longhi:15} and
was subsequently implemented experimentally in the work of~\cite{malomed},
albeit in the linear regime so far.
On the other hand, the experimental breakthroughs allowing careful control of the order of the dispersion derivative, notably for
integer (and even, in particular) orders, have sparked a renewed theoretical interest in this problem~\cite{TamOL2019,TamPRA2020,BandaraPRA2021,aceves} after initial theoretical interest some decades earlier~\cite{KarlssonOC1994,AkhmedievOC1994,KarpmanPLA1994}.  While the majority of investigations have focused on the appearance of bright solitary waves in these controlled dispersion-order systems, work in Kerr nonlinear microresonators~\cite{BaoJOSAB2017,TaheriOL2019,ParraRivasOL2016,ParraRivasPRA2016} suggests possible pathways for experimental realization of systems with dark solitary waves.  Theoretical/numerical investigations into the features of dark solitary waves in the presence of higher-order dispersion have uncovered a wealth of possible stationary states~\cite{Alexander:22}, stability features and self-similar nonlinear dynamics~\cite{DeckerJPA2020,DeckerCNSNS2021,TsoliasJPA2021,TSOLIAS2023107362}.  Dark-like modes have also been found in the fractional nonlinear Schr{\"o}dinger equation in the presence of a Gaussian potential barrier, however at fractional orders below the harmonic case~\cite{ZengND2021}.


The complex dynamics in the quartic derivative case, more specifically, arises due to to the oscillatory tails in the solitary wave structures, emerging due to the biharmonic operator.  Indeed, as was shown, e.g., in~\cite{DeckerJPA2020}, the nature of
the tails and that of the inter-soliton
interactions in the pure-quartic problem
involves oscillatorily-decaying
exponentials (i.e., a complex spatial 
eigenvalue). This leads to numerous equilibria
(of alternating stability) and intriguing
self-similar phase portraits. On the other hand,
it is a well-known feature of regular 
(harmonic) nonlinear Schr{\"o}dinger (NLS) solitons~\cite{KIVSHAR1995353}
---see also~\cite{pavloff}---, and of the
corresponding double-well ($\phi^4$) real
field theory~\cite{p4book} that no 
solitonic bound-states exist. Indeed, at
the level of the real $\phi^4$ field theory
a kink and an antikink attract, while for 
NLS dark solitons, it is well-known that they
repel~\cite{KIVSHAR1995353}. 

Naturally, this suggests a disparity.
On the one hand, 
there is the behavior of the harmonic model
which has only real spatial eigenvalues,
exponentially interacting structures and
no bound states. On the other hand,
the biharmonic model features complex
spatial eigenvalues, oscillatorily 
exponential interactions and an infinity
of bound states. It is then natural
to ask: if one could afford to have a 
continuous {\it fractional derivative}
parameter that would interpolate between
the two limits, how would the relevant transition
occur? It is this question that we
pose in the present work, seeking to 
address it at the level of the real
($\phi^4$) field theory. The latter
choice is made in order to avoid
the complications of the complex field
theory and its associated phase dynamics.

Our concrete toolbox will involve the
formulation of the $\phi^4$ (real) field-theoretic
problem at the level of the Riesz (spatial)
fractional derivative. In what follows,
we first provide a general theoretical
formulation of the problem setup in Section II.
We give a brief background for the
reader on different fractional derivatives and 
present our tool of choice, namely the
Riesz spatial fractional derivative.
Then in Section III, we study the tails of
a single kink, en route to the 
consideration of kink-antikink steady states
in Section IV. This contains also
the main result of our work regarding
how relevant bound states ``evaporate''
as one transitions from the biharmonic
to the harmonic case. Details of our
calculations as concerns the tail
dependence on the fractional exponent
and the kink-antikink force as a function
of the separation are provided in 
Sections V and VI, while the case of
values outside the harmonic to biharmonic
range (of exponents outside the interval
$[2,4]$) is briefly discussed in 
Section VII. Finally, Section VIII
summarizes our findings and presents our
conclusions.

\section{\protect\vspace{1pt}Problem Setup}

\subsection{PDE Models}
In our earlier works we have addressed equations of the form~\cite{TsoliasJPA2021}
\begin{equation}
u_{tt}=au_{xx}-bu_{xxxx}-u(u^{2}-\mu ),  \label{eq:beam}
\end{equation}%
i.e., a $\phi^4$ model with quadratic and quartic dispersion, and cubic nonlinearity,
and also NLS variants thereof~\cite{TSOLIAS2023107362,Alexander:22}
\begin{equation}
iu_{t}=au_{xx}-bu_{xxxx}-u|u|^{2}.  \label{eq:nls}
\end{equation}%
In the former case we first looked for solutions of the form $%
u(x,t)=\phi (x)$ (steady-state solutions) and in the latter case we 
explored standing wave solutions of the form $u(x,t)=e^{i\mu t}\phi (x)$. In both cases we end up (at the steady state level) with the equation%
\begin{equation}
a\phi ^{\prime \prime }-b\phi ^{(iv)}-\phi (\phi ^{2}-\mu )=0  \label{eq:stat}
\end{equation}%
assuming a real-valued $\phi $.

With $a=1$ and $b=0$ we get pure quadratic dispersion in each model, and
with $a=0$ and $b=1$ we get pure quartic dispersion. We get
\textquotedblleft in between\textquotedblright\ cases for various values of
nonzero $a$ and $b$. In this paper we use a different approach which ties
together the quartic and quadratic cases, giving a different  parametrization
between these two extreme cases using fractional derivatives. Letting $%
\alpha $ be the order of the derivative, we let $D^{\alpha }$ be the (Riesz)
fractional derivative operator of order $\alpha $. We use the Riesz
definition as it interpolates smoothly between our endpoint cases of $\alpha
=2$ ($a=1$, $b=0$) and $\alpha =4$ ($a=0$, $b=1$). In particular, it can be
shown that for $\alpha =2$ we have $D^{\alpha}f=f^{\prime \prime }$, and for $%
\alpha =4$ we have $D^{\alpha}f=-f^{(iv)}$. The negative sign for the $\alpha =4$
case is quite convenient, in that the $u_{xx}$ term enters with a positive
sign and the $u_{xxxx}$ terms enters with a negative sign in the equations
above, thus allowing for smooth interpolation as $\alpha $ ranges from $2$
to $4$. Another way to phrase this is that we preserve the
sign-definiteness of the two end points of the relevant interpolation.
Thus equations (\ref{eq:beam},\ref{eq:nls},\ref{eq:stat}) above can now be written as models encompassing the different special cases as respectively:
\begin{equation}
u_{tt}=D^{\alpha }u-u(u^{2}-\mu )  \label{beam1}
\end{equation}%
\begin{equation}
iu_{t}=D^{\alpha }u-u|u|^{2}  \label{nls1}
\end{equation}%
\begin{equation}
D^{\alpha }\phi -\phi (\phi ^{2}-\mu )=0  \label{steady1}
\end{equation}%
In this paper, and in order to understand the most prototypical
of the relevant settings,
we will make use of the real field-theoretic dynamical model of Equation (\ref{beam1}) and its stationary
analogue of Eq.~(\ref{steady1}%
) and leave Equation (\ref{nls1}) for future work, where potential
complex field-theoretic and phase-driven effects may come into play. Numerical methods will be 
used to find and continue steady states and to simulate the 
corresponding dynamics and the resulting features
will be theoretically interpreted and will be seen
to pose numerous interesting issues for future studies.


\subsection{The Riesz Derivative}
\label{sub:riesz}

A wide variety of fractional derivatives 
has been defined in the literature.  
The
most commonly used are the Riemann--Liouville, the
Caputo, and the Fourier~\cite{podlubny1999fractional,Samko}. See Appendix \ref{app:fracDef} for details. For fractional
boundary value problems (and consequently PDEs with fractional dispersion)
a type of average of the left and right derivatives is often preferred, 
namely the so-called
Riesz derivative~\cite{spaceFracQM}. We will denote the Riesz derivative of order $\alpha $ as $%
D^{\alpha }$.\newline

\emph{Riesz derivative (definition 1)}

$D^{\alpha }f=-\frac{D_{L}^{\alpha }f+D_{R}^{\alpha }f}{2\cos (\alpha \frac{%
\pi }{2})}$\newline

where we have suppressed the $x$ dependence. For odd integers this is not
well defined, but the limit as $\alpha \rightarrow 1,3,5,...$ does exist.
Our second definition for the Riesz avoids this problem, and can be shown to
agree with the first definition when $\alpha $ is not an odd integer.\newline

$\vspace{1pt}$\emph{Riesz derivative (definition 2)}

$D^{\alpha }f=-F^{-1}[(|k|^{\alpha })\widehat{f}(k)].$

$\vspace{1pt}$

Here, the hat symbol denotes the FOurier transform
and $F^{-1}$ its inverse.
Based on the above, definition 2 is the one used in our numerics.

$\vspace{1pt}$

\begin{figure}[]
\begin{center}
\includegraphics[width=5.5cm]{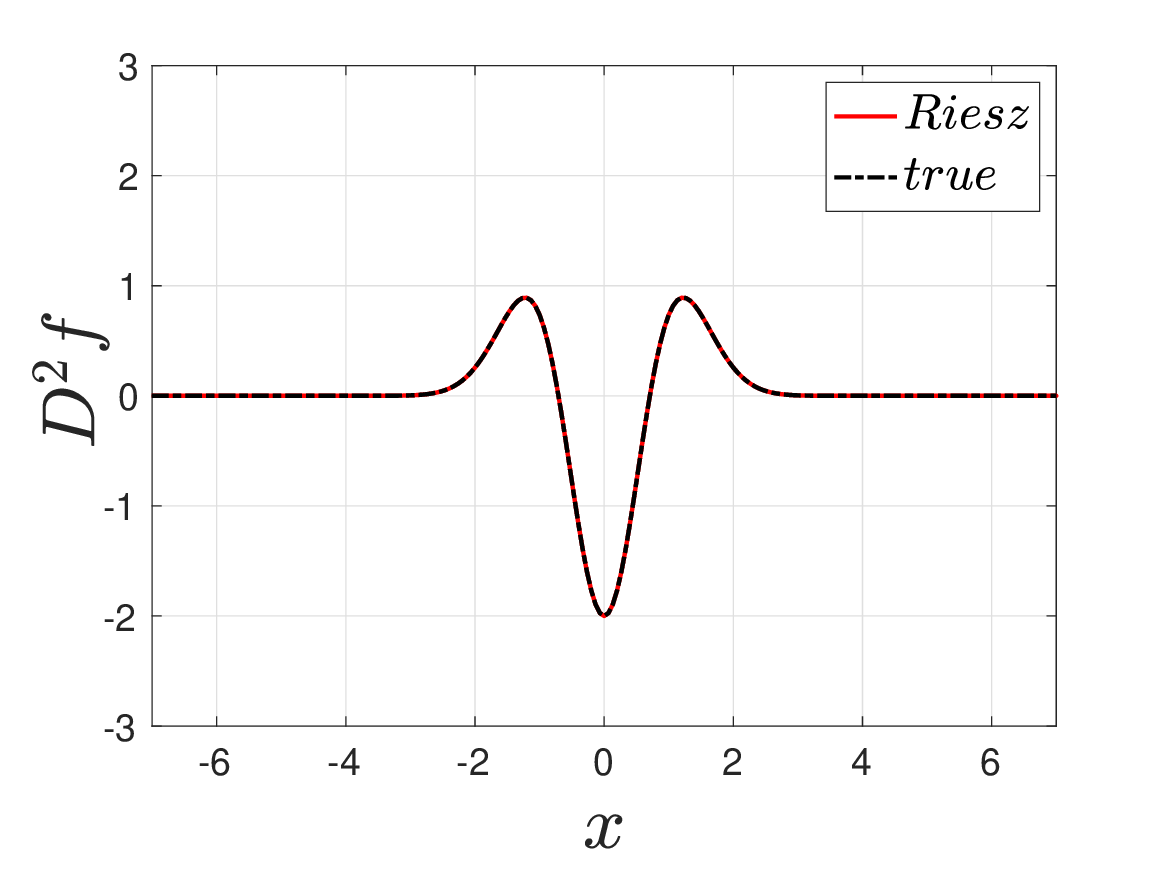}
\includegraphics[width=5.5cm]{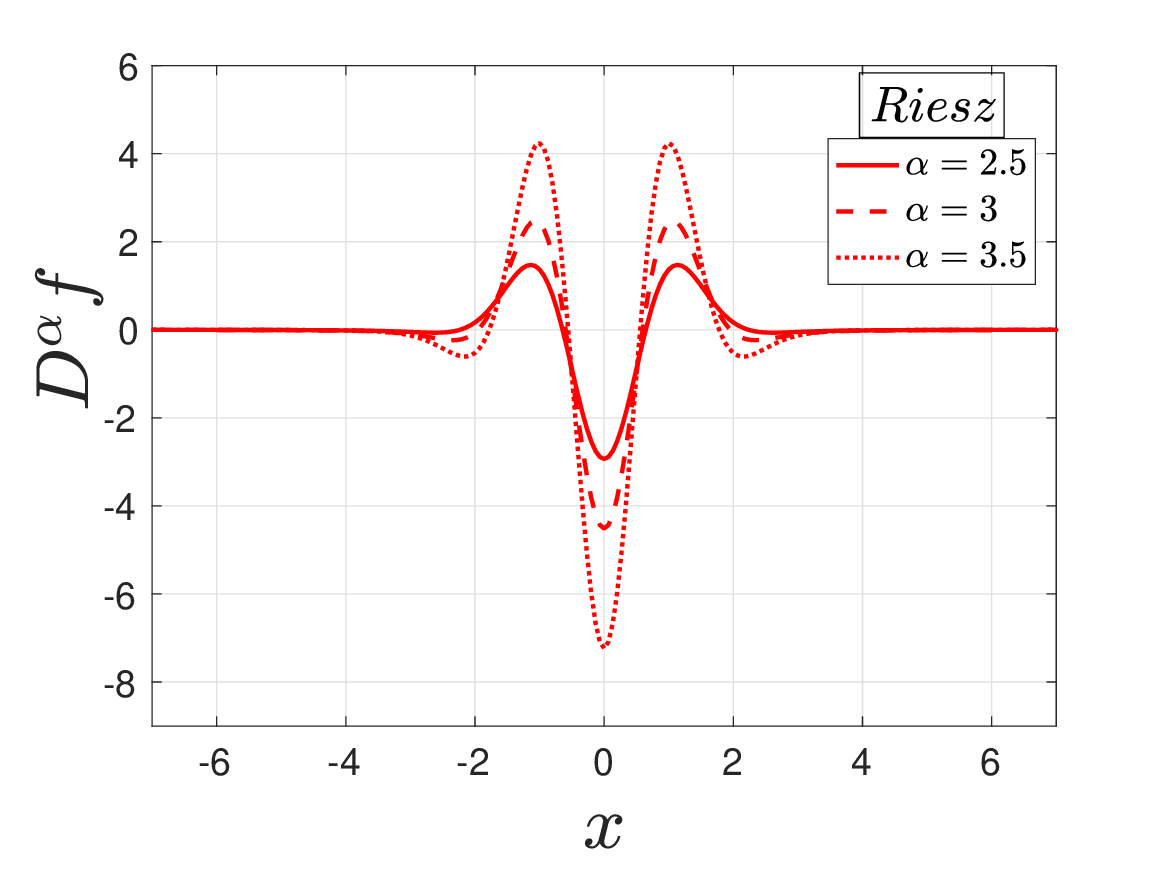}
\includegraphics[width=5.5cm]{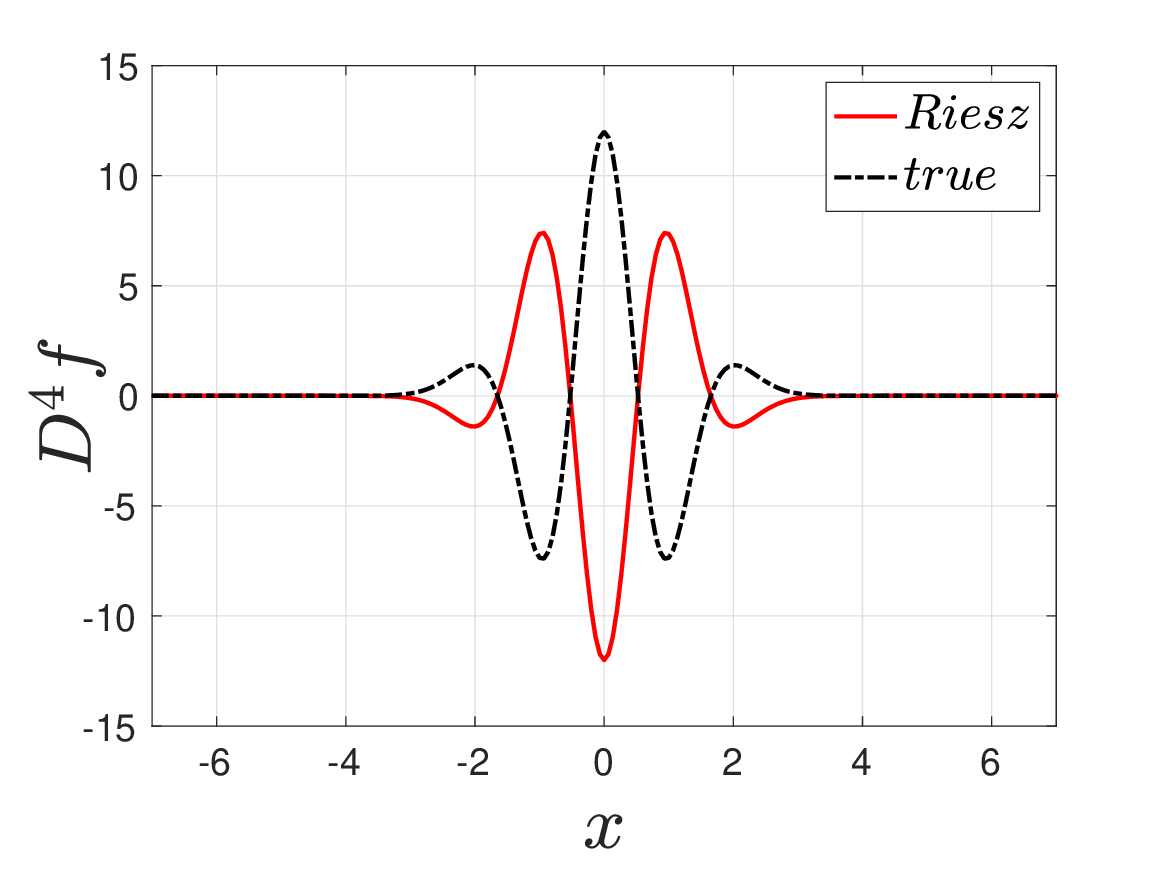} %

\end{center}
\caption{Left and right panels present the Riesz derivative (red solid) of $f(x)=e^{-x^2}$ of order $\alpha$, for $\alpha=2$ and  $\alpha=4$ along with standard second and fourth derivative (black dash-dot) respectively. The middle panel shows the  Riesz derivative of $f(x)=e^{-x^2}$ of order $\alpha = 2.5$ (red solid), $\alpha = 3$ (red dash) and $\alpha = 3.5$ (red dot).}
\label{alpha24}
\end{figure}

In Figure \ref{alpha24} we show Riesz derivatives for $\alpha=2$ through $\alpha=4$ in 0.5 increments. Note that the general shape is the same for each $\alpha$ and that the Riesz derivative at $\alpha=2$ agrees with the standard derivative and the Riesz derivative at $\alpha=4$ agrees with the negative of the standard derivative there. In between, the Riesz derivative smoothly interpolates between these extremes. The Gaussian is a reasonable example for our problem, as it resembles a kink-antikink interaction for small separation. Also, see Figure \ref{alpha3} (in Appendix \ref{app:fracDef}) for a comparison of left,  right and Riesz derivatives.

\section{Numerical Methods}\label{numericalMethods}

\subsection{Steady States and Numerical Continuation}\label{steadyStates}
We can discretize Eq. (\ref{steady1}) by first discretizing the continuous $x$ variable on an interval $[-L,L]$ using $N+1$ equally spaced intervals to get the discrete $x_i$. Base values of at least L=80 and N=1200 were used, with larger values required in some cases. Then with $\phi_i=\phi(x_i)$ we can approximate the Riesz derivative of order $\alpha$ using the Fourier definition as described in Section \ref{sub:riesz}, definition 2, with the FFT replacing the Fourier Transform.

Steady states can then be found using the Matlab `fsolve' command on the left side of the discrete version of Eq. (\ref{steady1}), along with a suitable initializer. The single kink initializers for all $\alpha$ were the easily verified solution for the case $\alpha=2$ given by $\phi_{0}(x)=\tanh(x/\sqrt{2})$ for our chosen $\mu=1$. For kink-antikink pairs we can use the function
\begin{equation}
\phi(x) = \phi_{0}(x+\delta)-\phi_{0}(x-\delta)-1,\label{KAKinitial}
\end{equation}%
as an initial guess for the stationary solution, where $2\delta$ is the separation distance between kink and antikink. For a (typical) given $\alpha$, there will be a finite number of values of $\delta$ which result in steady states. We can find those $\delta$ values and the corresponding steady states using a simple numerical continuation, starting at $\alpha=4$ (see Section \ref{numCon}) and then reducing $\alpha$ by a small increment and using the previous steady state as the initializer for the next steady state. The corresponding $\delta$ is determined by intersecting the new steady state with $\phi=0$ (positive solution). 

To determine stability of the various steady states, we created a fractional differentiation matrix
$D^\alpha$, again based on definition 2 of Section \ref{sub:riesz}, with discrete Fourier Transformation matrices replacing the continuous operators. Stability can then be determined by using the Matlab `eig' command to find the eigenvalues of
$\begin{bmatrix}
0 & I \\ 
A & 0%
\end{bmatrix}$
where $A=D^\alpha+I-3\phi^2$, and $\phi$ is the steady state.

\subsection{Zero Velocity States}\label{zeroVelocityStates}
We also need to create kink-antikink states where the kink and antikink are initialized to zero initial velocity, though they are not steady states. 
This is done so that we can measure the force
that the kink exerts on the antikink for {\it arbitrary}
values of $\delta$ and from that extract the corresponding potential of the interaction in what
is shown below.
This can be achieved by minimizing the norm of the left side of Equation (\ref{steady1}) using the initializer in Equation (\ref{KAKinitial}), subject to the constraint that the positions of both the kink and antikink stay fixed, as was initially proposed
in~\cite{Gani1}. For this we use the Matlab `lsqnonlin' command in the same way we use `fsolve' for steady states, applying it to the left side of the discretized version of Equation (\ref{steady1}). However, in this case we need to constrain the two nodes closest to $\phi=0$ which is done be setting both the upper and lower bounds for these two nodes to their values in the initializer (using lsqnonlin `lb' and `ub' parameters).

Practically, such states are used in Section \ref{sec:accel} to find acceleration as a function of kink-antikink separation by using them as initial conditions in a dynamic simulation and then finding the slope of the velocity curve over a short period of time. They are also used in dynamic simulations over a longer period of time to generate PDE phase portraits.

\subsection{Dynamic Simulations}\label{steadyStates}
For PDE simulations we discretize the right side of Eq. (\ref{beam1}) as before, resulting in a system of N ODEs, which can be numerically solved using the Matlab routine `ode45'. Initial conditions were zero velocity states as described above. We also use `ode45' to numerically solve the ODEs created from the acceleration data, as described in Section \ref{sec:accel}.

\section{Single Kink Tails}\label{singleKink1}

\begin{figure}[tbp]
\begin{center}
\includegraphics[scale=0.5]{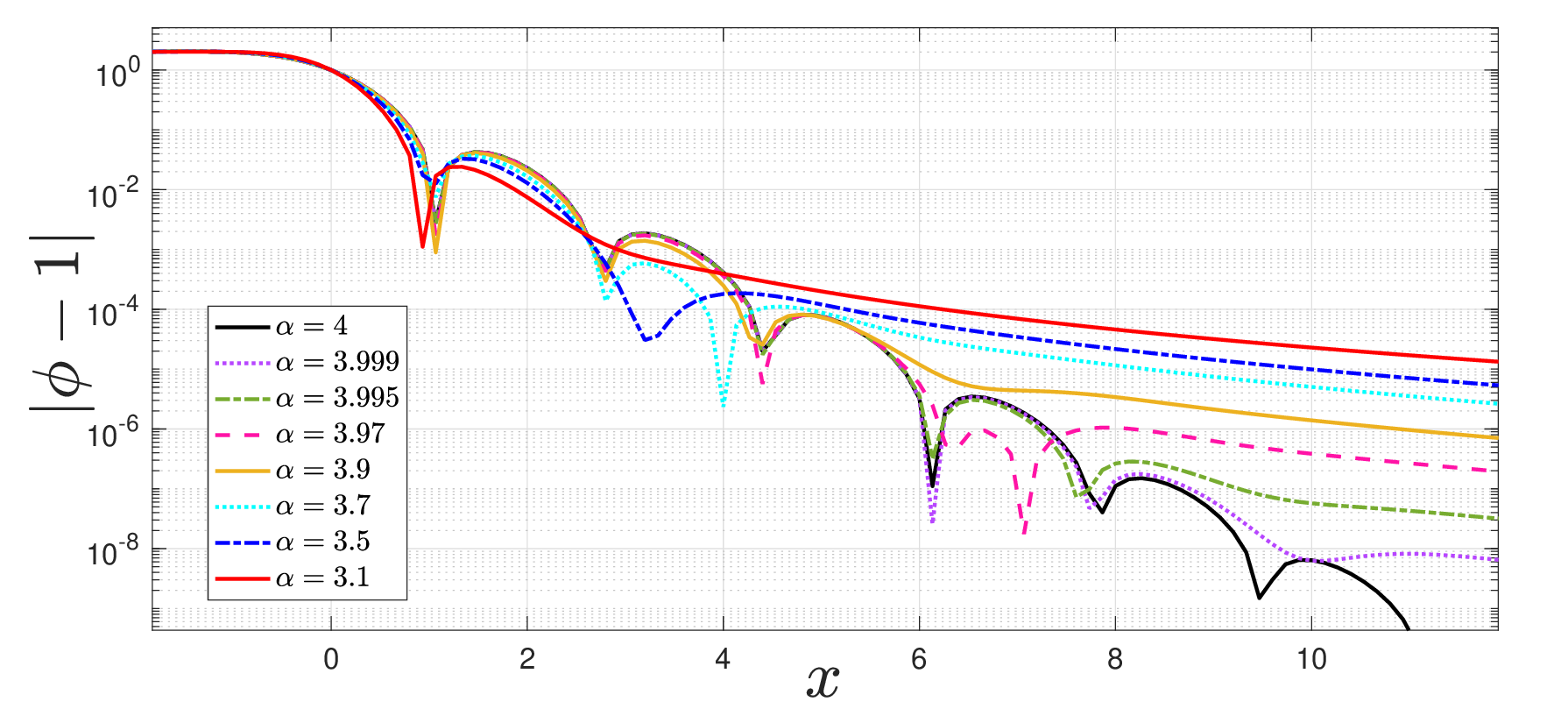}
\end{center}
\caption{Tails of a single kink for various $\alpha$ on a semilog-y plot,
showing the distance from the asymptotic value $|\phi-1|$ as a function
of the spatial variable $x$.}
\label{fig:tail_plots_13}
\end{figure}

We now turn to the single kink solution of the 
stationary Eq.~(\ref{steady1}).
In Figure \ref{fig:tail_plots_13} we plot $x$ on the horizontal axis and $|\phi-1|$ on the
vertical axis (semilog plot) for a single kink (with tails asymptotic to $%
\phi=\pm1$) for various values of the derivative order $\alpha$. One sees that as $\alpha$ increases from 2 to 4, the number of
times that the tail crosses $\phi=1$ increases. Each ``sharp'' minimum (where
the graph would actually go to $-\infty$, indicating a zero of $\phi-1$)
indicates a new crossing. 
That is, there is one such zero at $\alpha=3.1$, two at $\alpha=3.5$ etc.
Upon looking more closely at the tails of these
single kinks it appears that there is a combination of an inner
region featuring a complex exponential decay
(sin/cos with exponential decay envelope in the region of the tail where
crossings of the asymptote occur) and an outer region exhibiting a
power-law decay (in the region of
the tail where crossings no longer occur).

\begin{figure}[]
\begin{center}
\includegraphics[width=5.5cm]{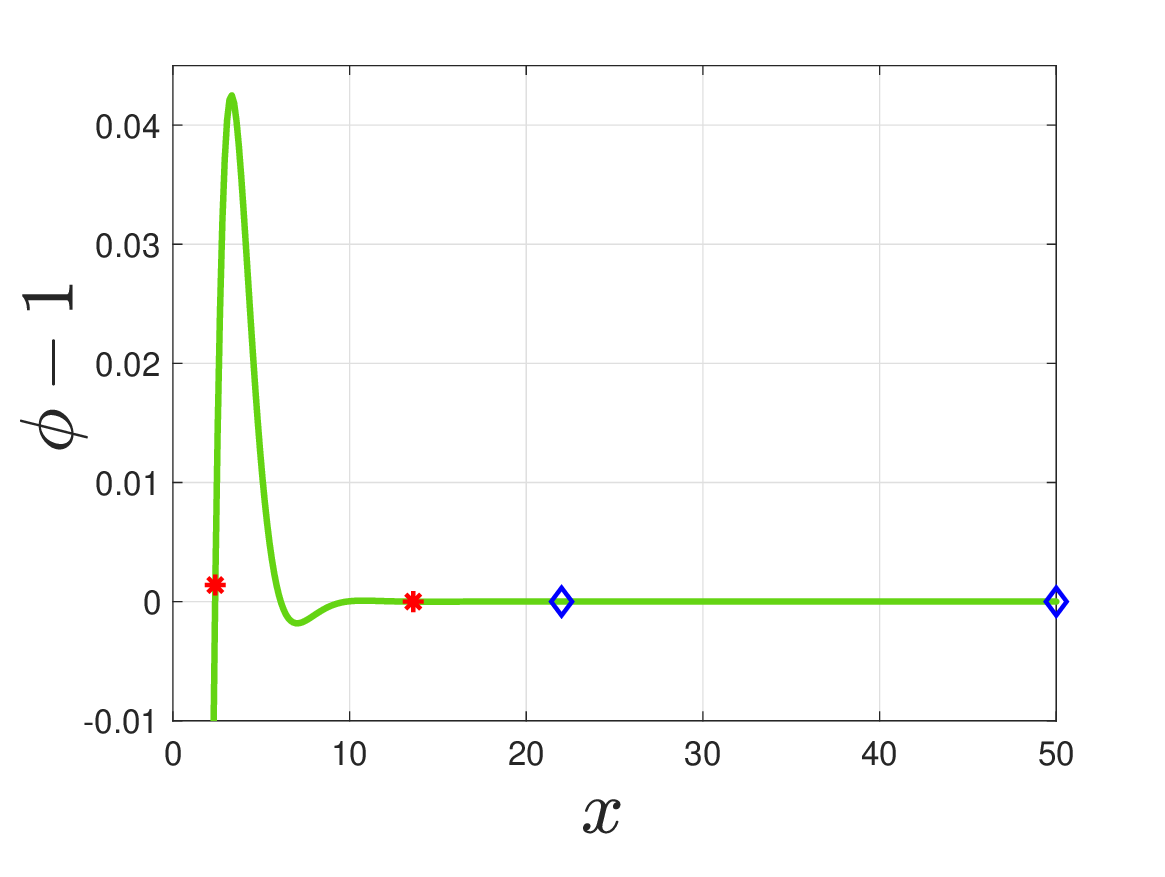}
\includegraphics[width=5.5cm]{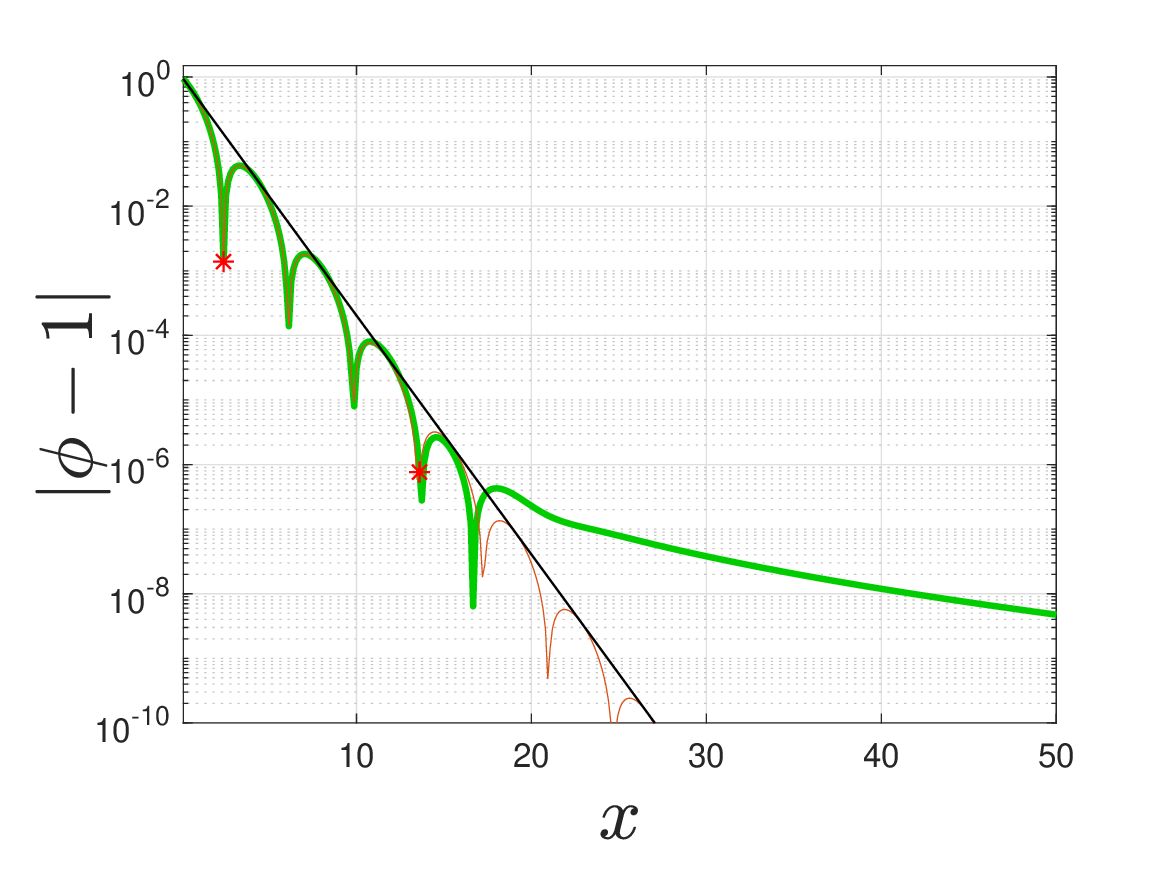} %
\includegraphics[width=5.5cm]{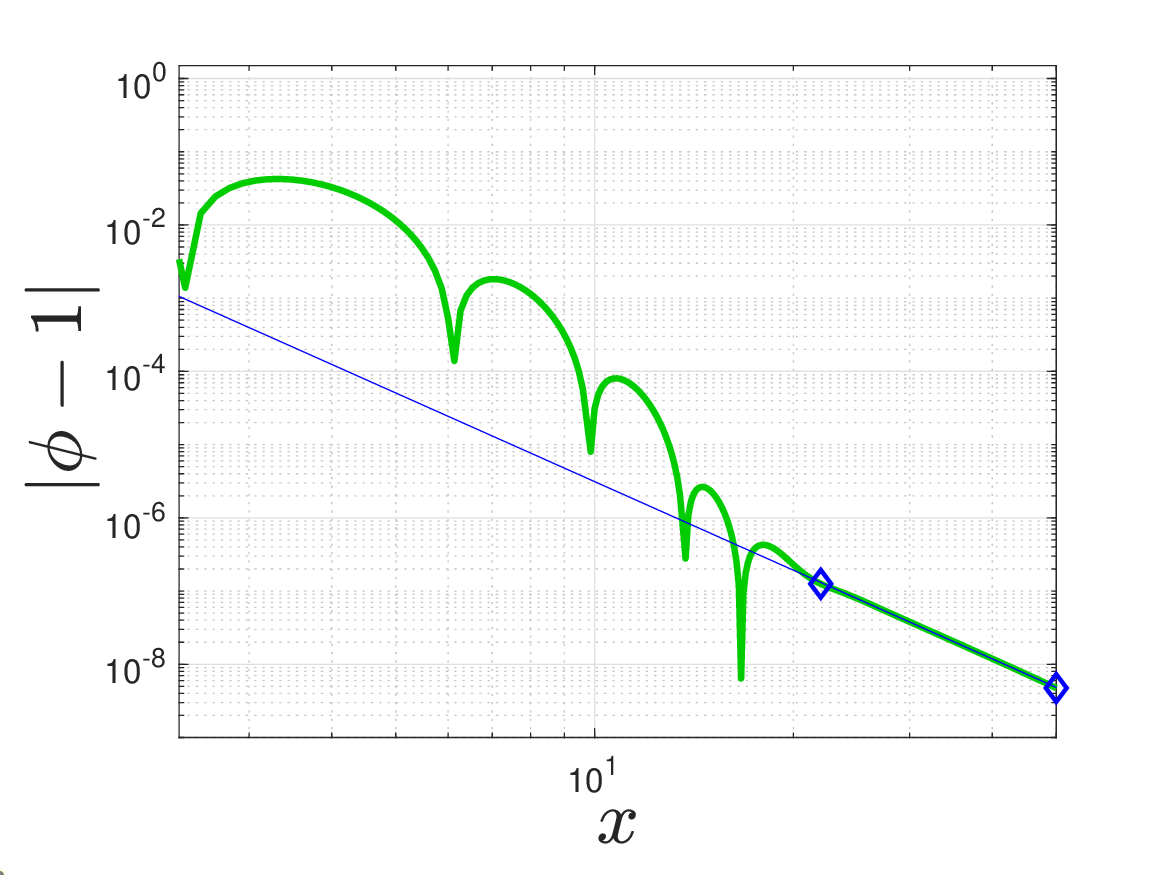}
\end{center}
\caption{Left panel: Kink tail for $\alpha=3.99$ (translated down one unit). A decaying cosine curve can be fit to the
region of the tail between the red asterisks and a power law 
curve can be fit
in the region of the tail between the blue diamonds. Middle panel: Kink tail (green curve) $|\phi-1|$ versus $x$ using semilog-y
plot. Also shown is a
decaying cosine curve ($-1.0128 \exp(-0.8517 x) \cos(0.8455 (x-7.9604))$ in
red) fitted to the green curve between the red asterisks, as well as its envelope
curve ($-1.0128 \exp(-0.8517 x)$ in black). Right panel: Kink tail (green curve) $|\phi-1|$ versus $x$ using loglog
plot. Also shown is a power
law curve ($0.02974 / x^{3.9907}$ in blue) fitted to the green curve between the
blue diamonds, accurately capturing the outer part of the tail.}
\label{tails2}
\end{figure}

The relevant details can more clearly be discerned in Figure \ref{tails2} where we show the tail (in green) for alpha=3.99, using three plot types.
The semilogy plot (Figure \ref{tails2} middle panel) shows a decaying cosine curve
(in red) fitted to the left part of the tail (portion between red
asterisks in the tail plot of the left panel). The envelope curve for the fitted portion (straight line) is shown
in black. 
Clearly, such a fit is very accurate in the inner region of the 
tail between the red asterisks.
The log-log plot (Figure \ref{tails2} right panel) shows a straight line in the
right of the tail (portion between the blue diamonds
in the tail plot of the left panel), and a power law curve,
fitted to the points in the tail between the blue diamonds, verifies that
this part of the tail is nearly perfectly straight, i.e., perfectly
conforms to a power-law decay.








\section{Kink-antikink steady states and numerical continuation/bifurcation} \label{numCon}

\begin{figure}[]
\begin{center}
\includegraphics[width=8cm]{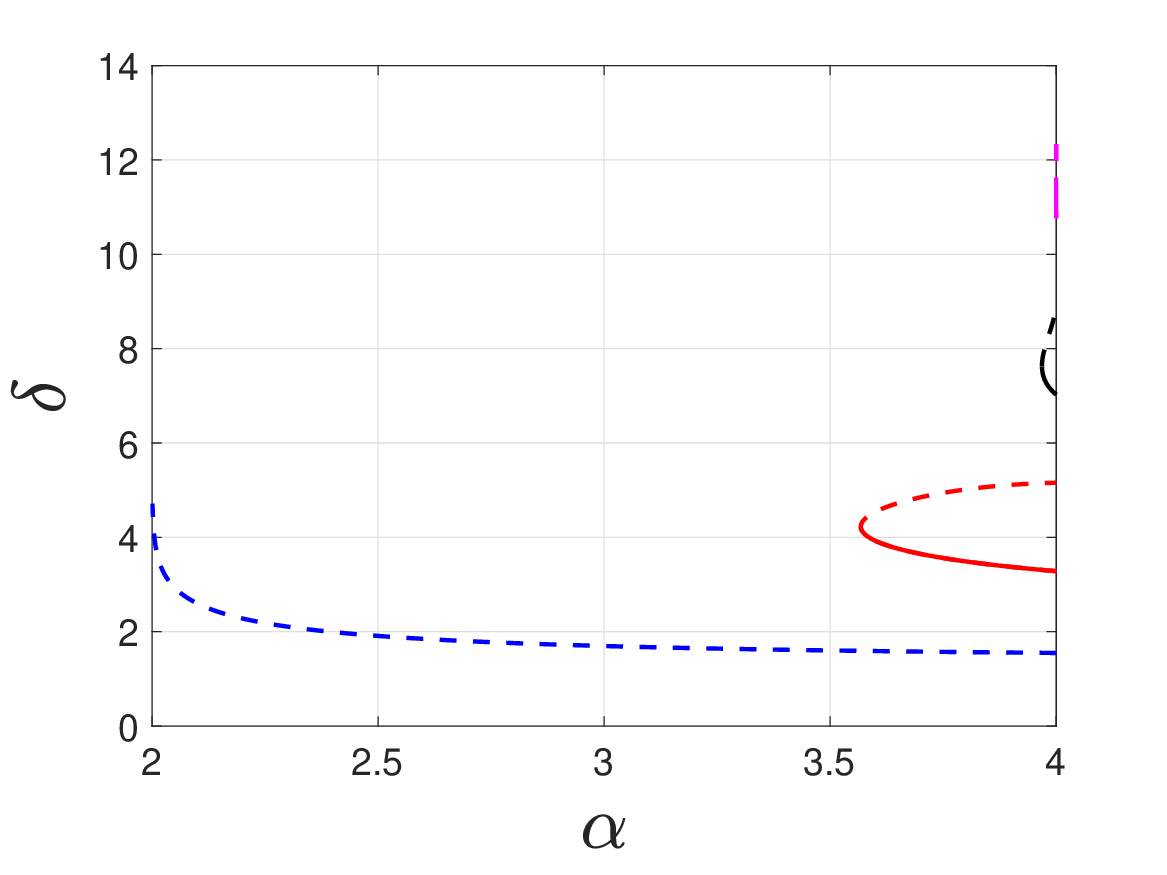}
\includegraphics[width=8cm]{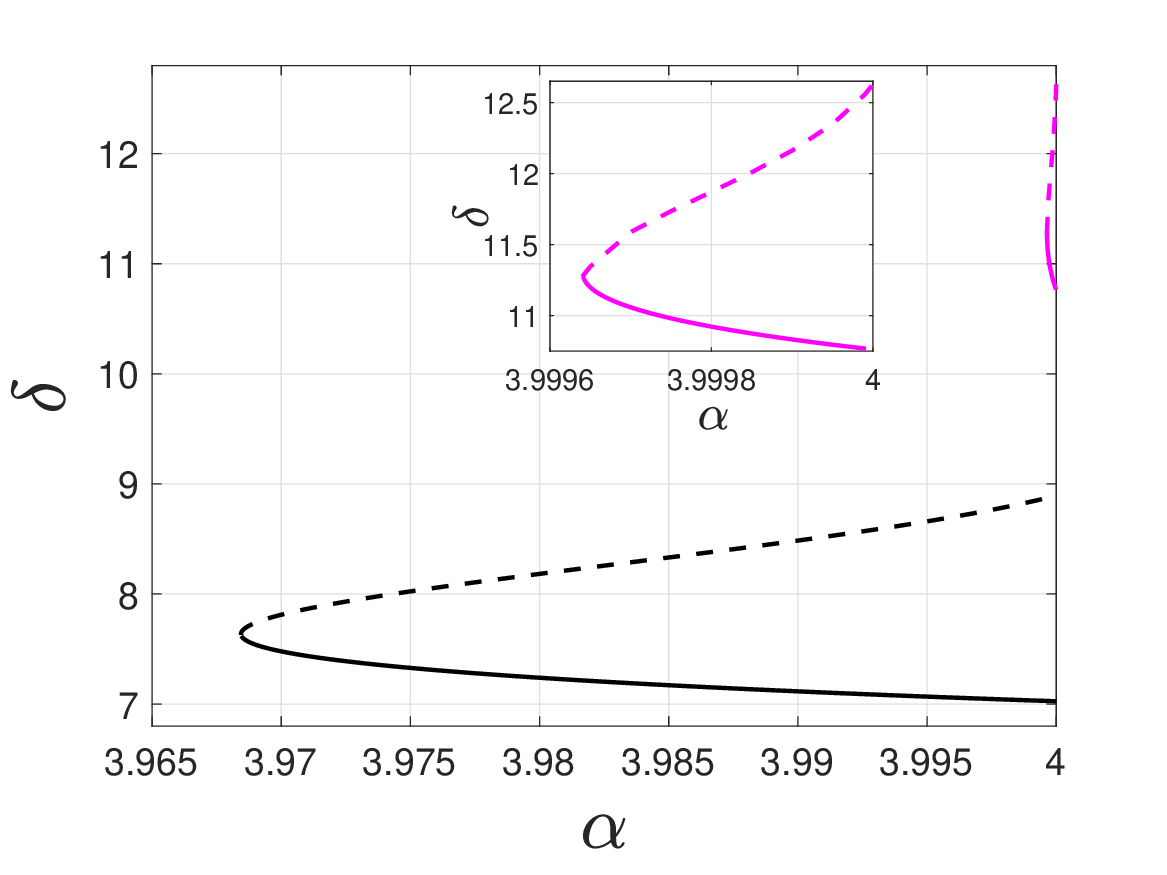}
\end{center}
\caption{Numerical continuation diagram for K-AK separation vs $\alpha$ with two zooms in the right
panel.  The vertical axis is the position of the antikink $\delta$, which is the half-separation value of kink and antikink. The first few $\alpha$ bifurcation points, where new steady states appear for sufficiently large separation values, are approximately $\alpha=3.5677, 3.9684, 3.99964$.
These critical points approach the biharmonic limit 
exponentially rapidly as the order of the saddle-center bifurcation 
increases.}
\label{bifurcate1}
\end{figure}

In an earlier work~\cite{TSOLIAS2023107362}, we found steady-state solutions for kink-antikink pairs using Equation (\ref{steady1}) with $\alpha=4$ for the six smallest kink-antikink separation values, labeling them as states 0-5. In that case there were infinitely many steady states. In this paper we find steady states of kink-antikink pairs as a function of $\alpha$ which becomes now the relevant {\it bifurcation parameter} for our continuation between the biharmonic and the harmonic case. New steady states appear (in pairs) as $\alpha$ approaches 4, or vice-versa
these states disappear pairwise as one departs from the biharmonic
limit. Each new pair consists of one stable and one unstable steady state, i.e., a center and a saddle state, and are connected by a numerical continuation, i.e., the relevant states terminate
in {\it saddle-center bifurcations} (as one departs from the biharmonic
limit, or vice-versa get born through such bifurcations as one increases
$\alpha$). The exception is the first state (state 0) for which the separation goes to infinity as $\alpha$ goes to two. 
Indeed, this state (of minimal separation in the biharmonic case)
appears to emerge through a {\it bifurcation from infinity} as soon as we depart from the harmonic
limit.
Figure \ref{bifurcate1} 
summarizes these pairwise bifurcations of states 1-2
(at $\alpha=3.5677$), 3-4 (at $\alpha=3.9684$), 5-6 (at $\alpha=3.99964$) and
so on, and the emergence from the infinite separation harmonic limit
of state 0 and represents the main result of the present work.
Notice the exponentially fast approach of the critical points
to the biharmonic limit
and also  that in the relevant bifurcation diagram 
the vertical axis is the position of the antikink, which is also the half-separation value of these symmetric states.
Also, as is commonly the case, the stability of each steady state is indicated by dotted lines for unstable states and solid line for stable ones (saddles and centers, respectively). 

\begin{figure}[]
\begin{center}
\includegraphics[width=5.5cm]{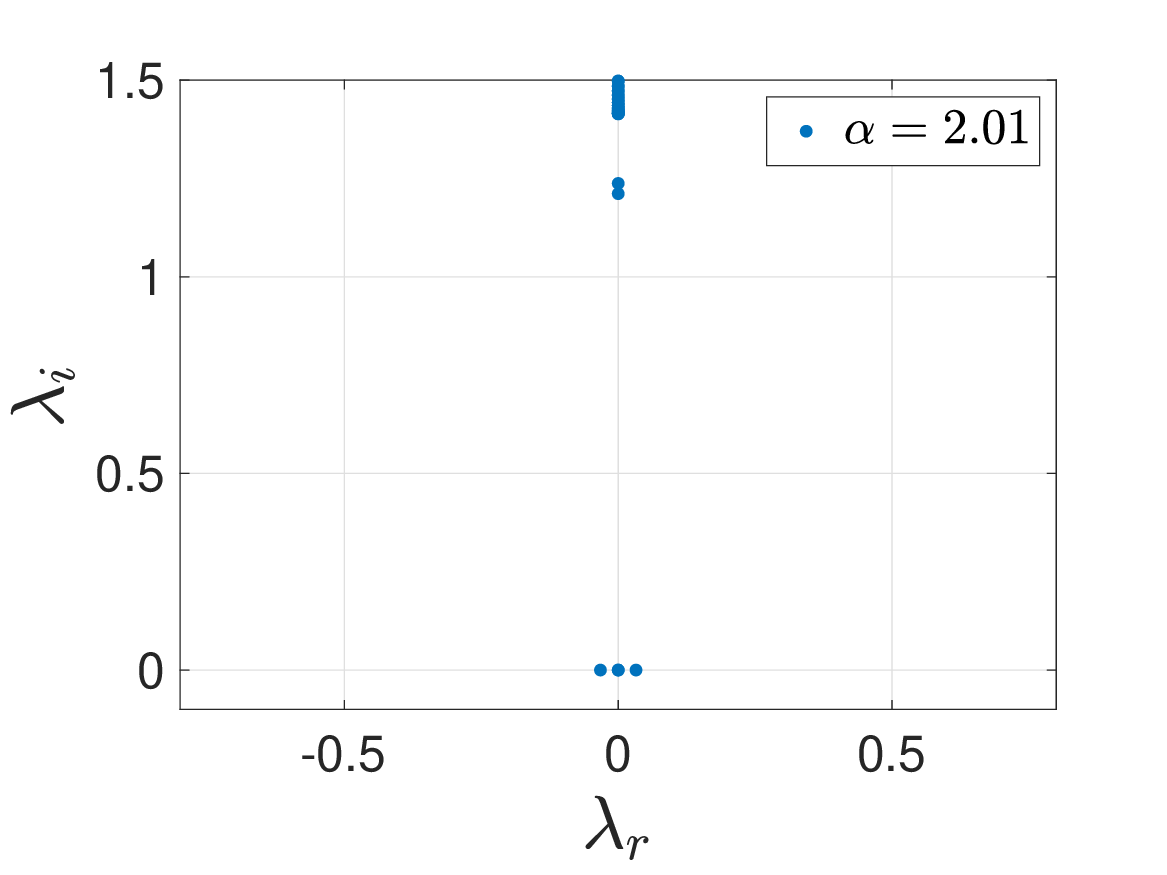}
\includegraphics[width=5.5cm]{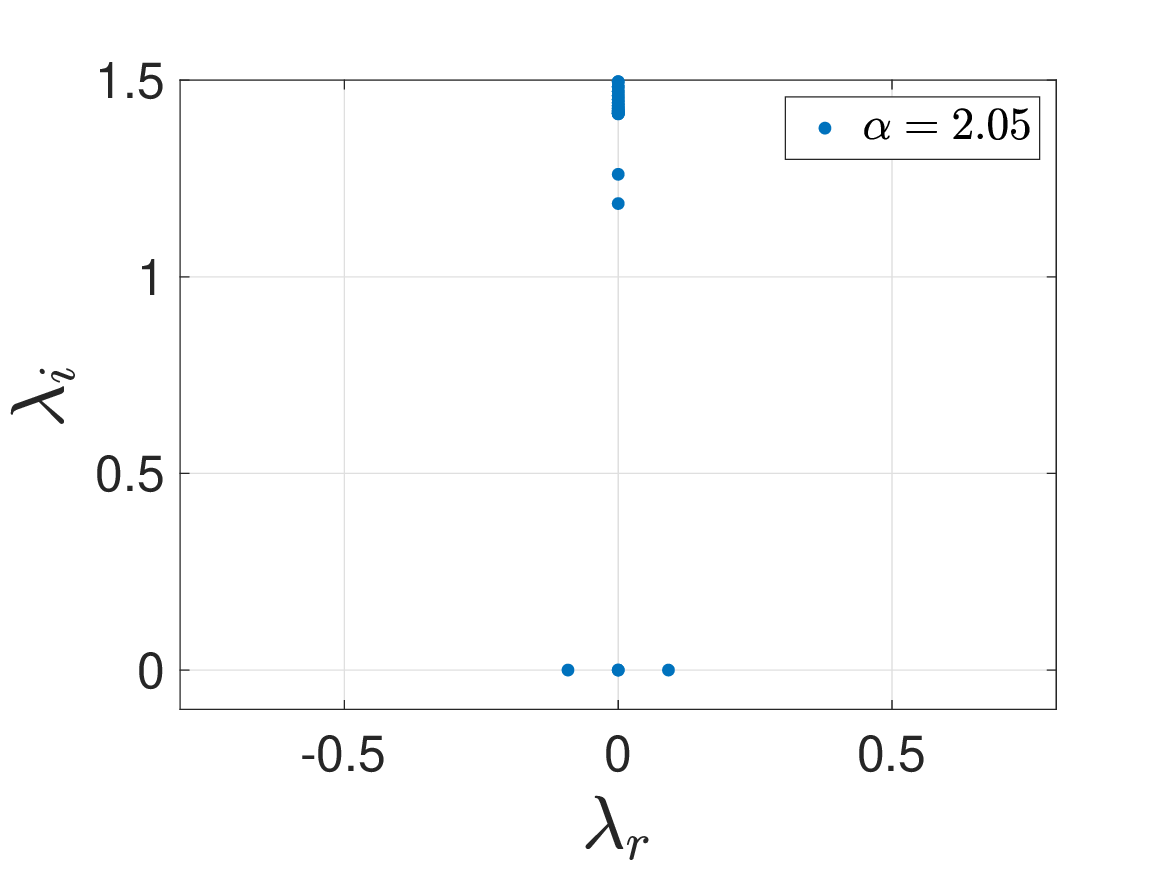}
\includegraphics[width=5.5cm]{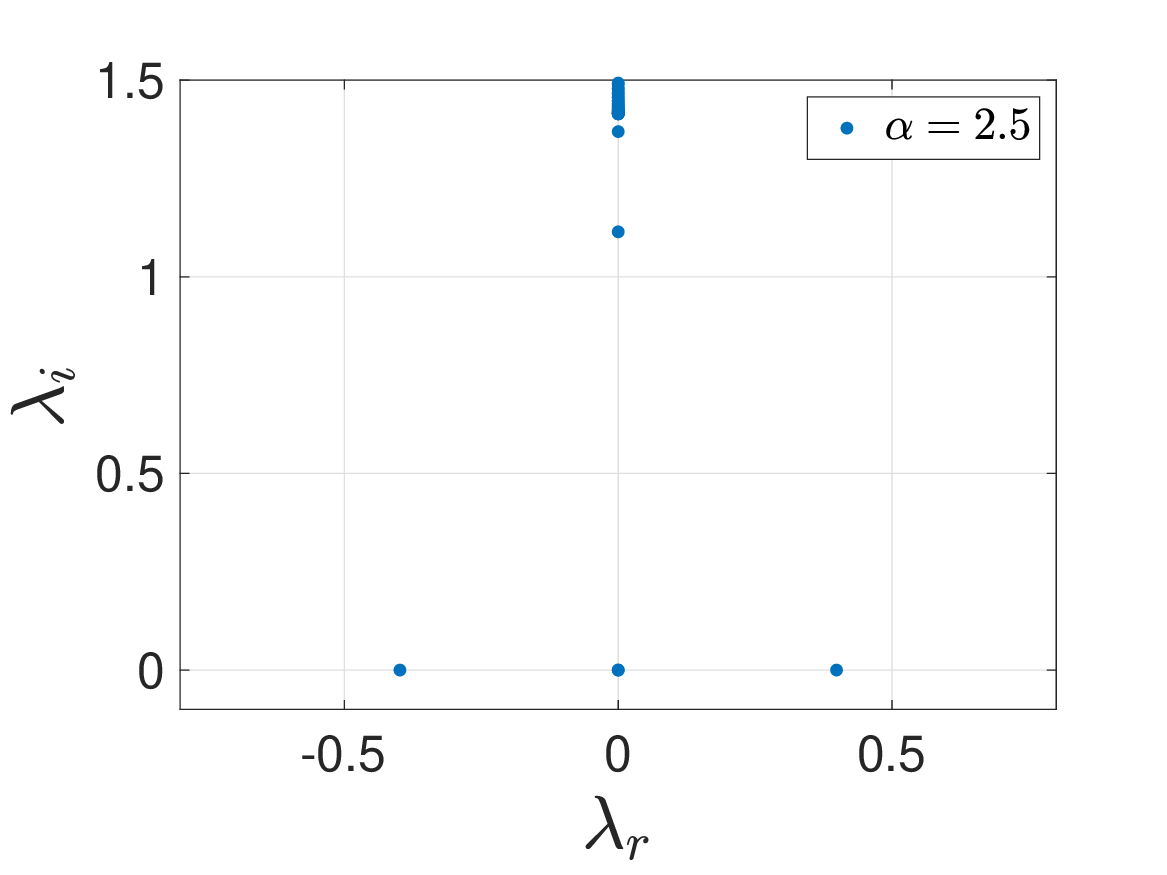}
\includegraphics[width=5.5cm]{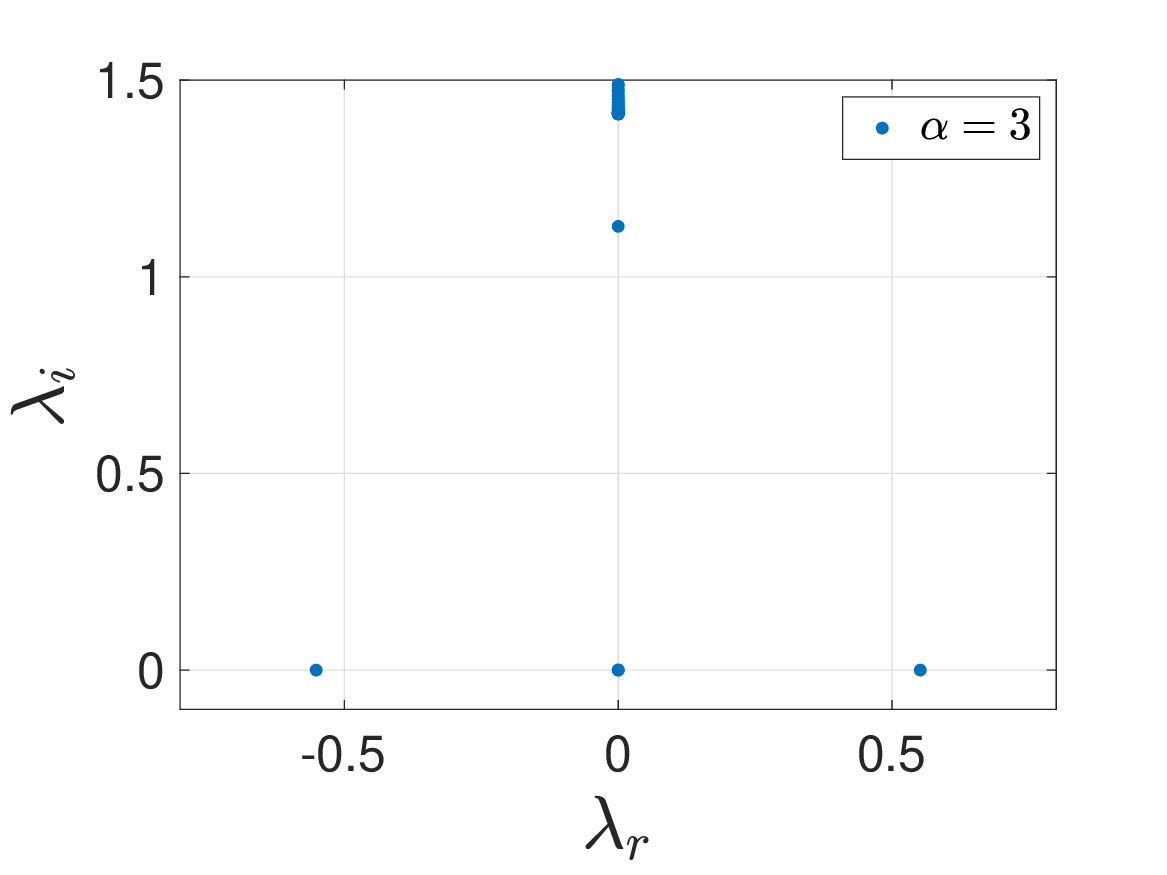}
\includegraphics[width=5.5cm]{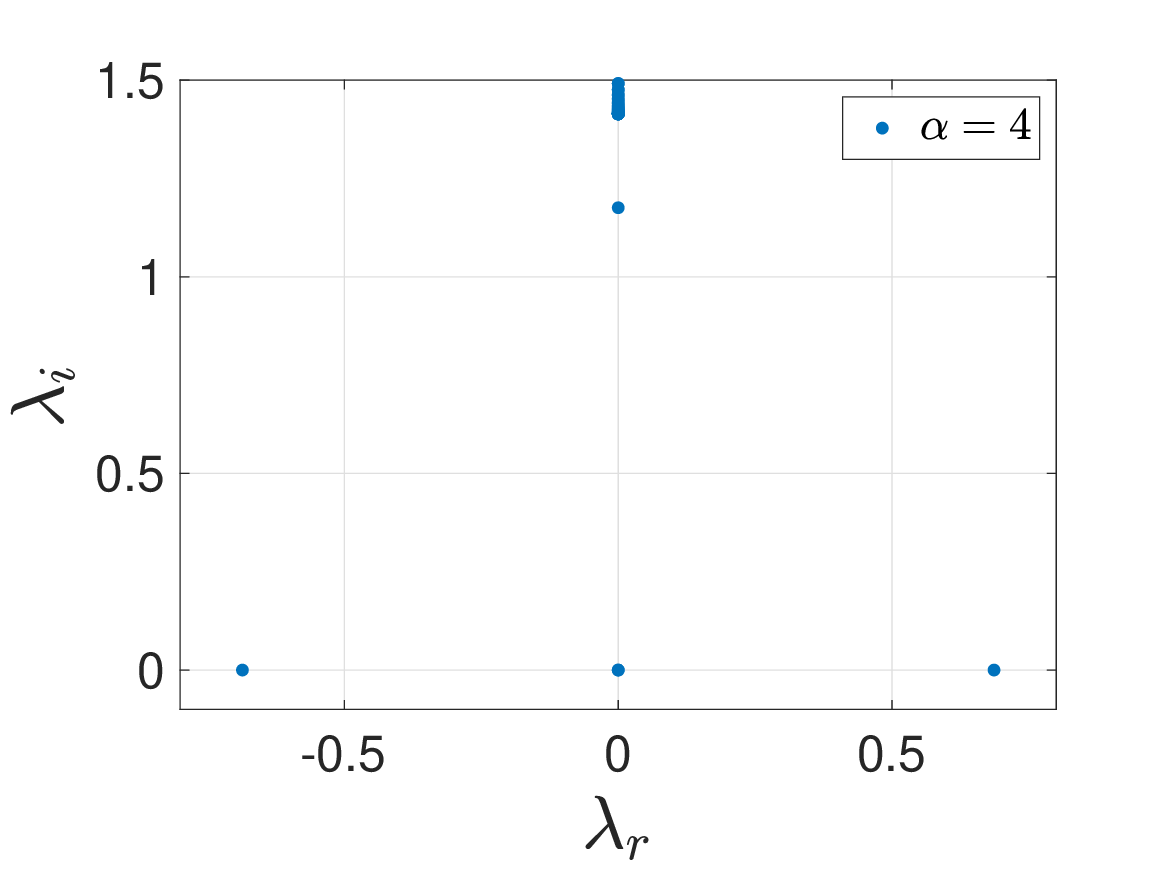}
\end{center}
\caption{Spectral plane $(\lambda_r,\lambda_i)$ plots of
the eigenvalues $\lambda=\lambda_r+i \lambda_i$  for steady states along the blue line in Figure \ref{bifurcate1}. From upper left to lower right, $\alpha$ values are 2.01, 2.05, 2.5, 3, 4. All plots indicate spectral instability. Furthermore, as $\alpha$ approaches 2, the spectral plots approach that of two indefinitely separated individual kinks.
Also note  how the pair of eigenvalues just below the continuous part of the spectrum ---starting at $\lambda= \pm \sqrt{2} i$--- splits, with the larger eigenvalue moving up into the continuous spectrum between the $\alpha$ values of 2.5 and 3 (see also the 
discussion in the text).}
\label{spectrum1}
\end{figure}

\begin{figure}[]
\begin{center}
\includegraphics[width=4.4cm]{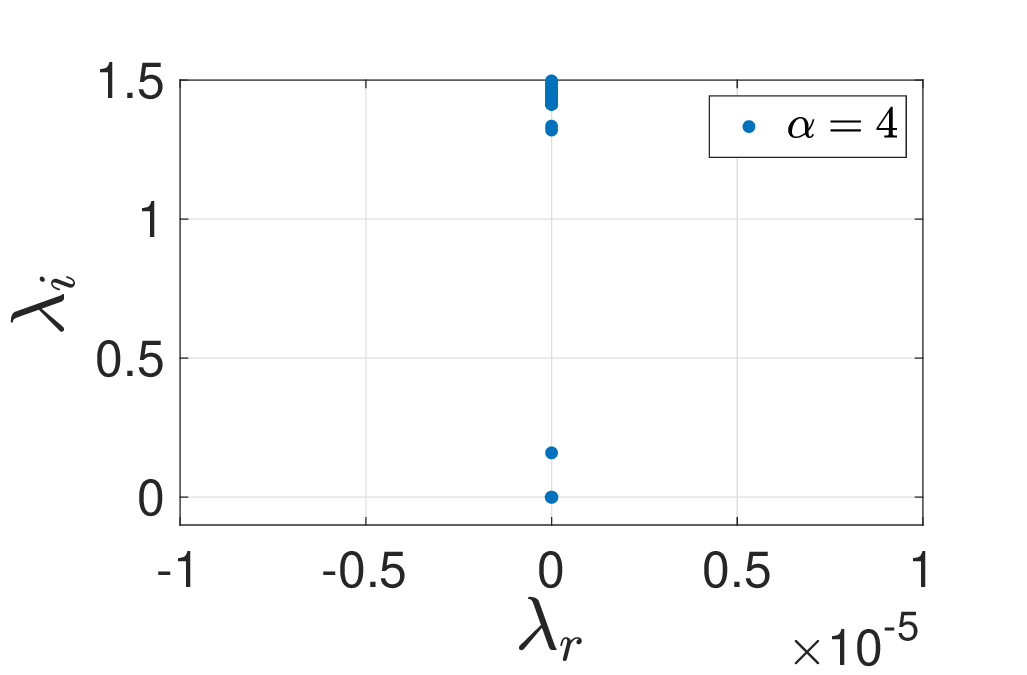}
\includegraphics[width=4.4cm]{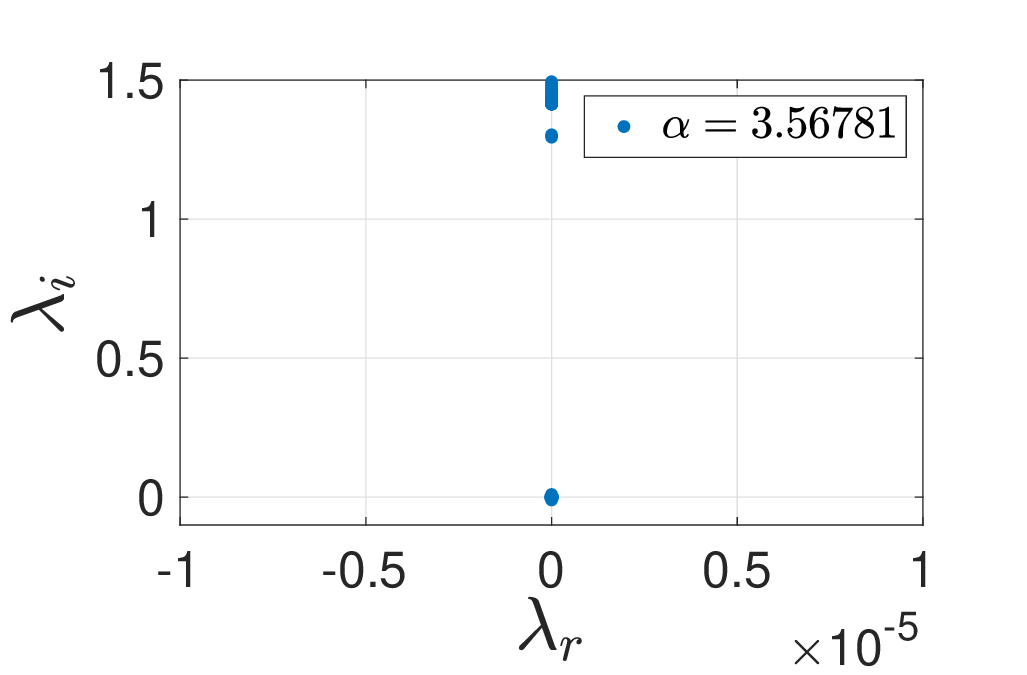}
\includegraphics[width=4.4cm] {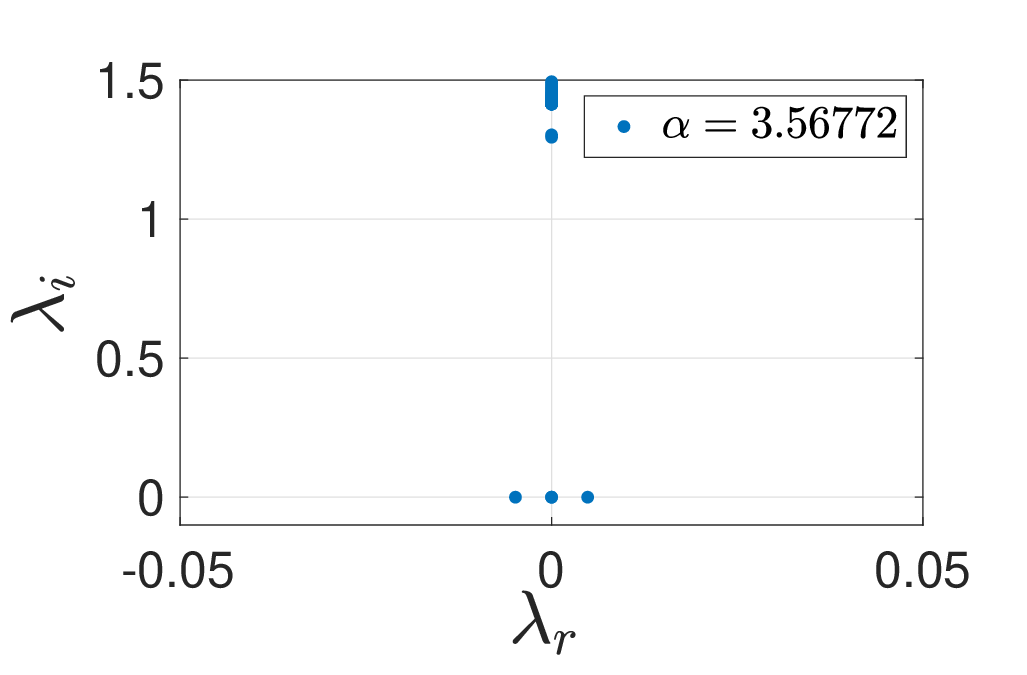}
\includegraphics[width=4.4cm]{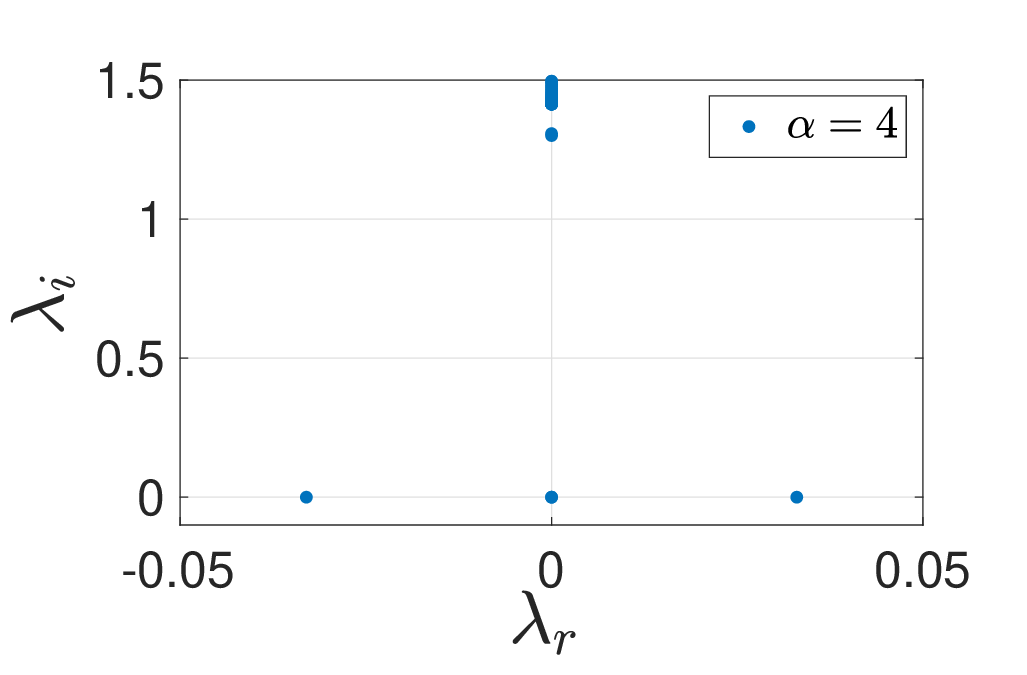}
\end{center}
\caption{Spectral plots for steady states along the red curve in Figure \ref{bifurcate1}. The order of the plots (left to right) corresponds to moving along the solid (stable) red line starting at $\alpha=4$ and moving clockwise around the red loop ending at $\alpha=4$ on the dotted (unstable) line. The two middle plots correspond to points very close to the turning point ($\alpha=3.56781$ on the stable line and $\alpha=3.56772$ on the unstable line) indicating the switch from stable to unstable.}
\label{spectrum2}
\end{figure}

In Figures \ref{spectrum1} and \ref{spectrum2} we show,
additionally, the eigenvalue plots for some steady states along the blue zeroth
branch (the one bifurcating from infinity at the harmonic limit)
and the red curve connecting the branches 1-2. In all cases only the positive y-axis is shown in the spectral plots, as the negative y-axis
involves a reflection thereof (without additional information).
It is interesting to observe that the blue branch is always unstable
with a growth rate that increases as the solitary waves
separation decreases. This is natural to expect as
in the harmonic limit of near-infinite separation, there
exist the in-phase translational mode (of 0 eigenvalue due
to the corresponding symmetry/invariance), as well as the out-of-phase
mode of instability. As $\alpha$ increases, the separation decreases,
increasing the interaction and hence the instability 
growth rate of this
out-of-phase motion. In Fig.~\ref{spectrum1}, there is a similar
effect on the celebrated kink internal mode~\cite{p4book}.
For $\alpha=2.01$, the relevant internal mode frequencies are
nearly indiscernible, given the large kink separation. As the
kink-antikink pair decreases its distance, the internal modes
involving breathing in- and out-of-phase become more distinct
in their respective frequencies. These types of features
(although with a lesser variation due to their more
limited interval of existence) can be identified also
in the case of the stable (left two panels, at 
$\alpha=4$ and just before the turning point)
and unstable (right two panels, right after
the turning point and at $\alpha=4$) red curve segments
of Fig.~\ref{spectrum2}.

\section{Dependence of single kink and kink-antikink tails on derivative order $\boldsymbol{\alpha}$}
In Section \ref{singleKink1} we saw that the tails of a single kink decay according to a power law as $x \rightarrow \infty$. In fact, extensive numerical investigation suggests that for a given value of $\alpha$ this power law is of the form ${\beta}/{x^\alpha}$ (where $\alpha$ is the order of the derivative) for all $2<\alpha<4$. Using a large interval ($-960<x<960$) in order to minimize boundary condition effects, we find that for $\alpha=2.01, 2.1, 3.0, 3.7, 3.99$ the slopes of the corresponding log-log curves are $2.0134, 2.1028, 3.0009, 3.7007 ,3.9907$. This close correspondence is strong numerical evidence that the tails obey a law of the form ${\beta}/{x^a}$ where $a$ is in fact $\alpha$.

\begin{figure}[]
\begin{center}
\includegraphics[width=12cm]{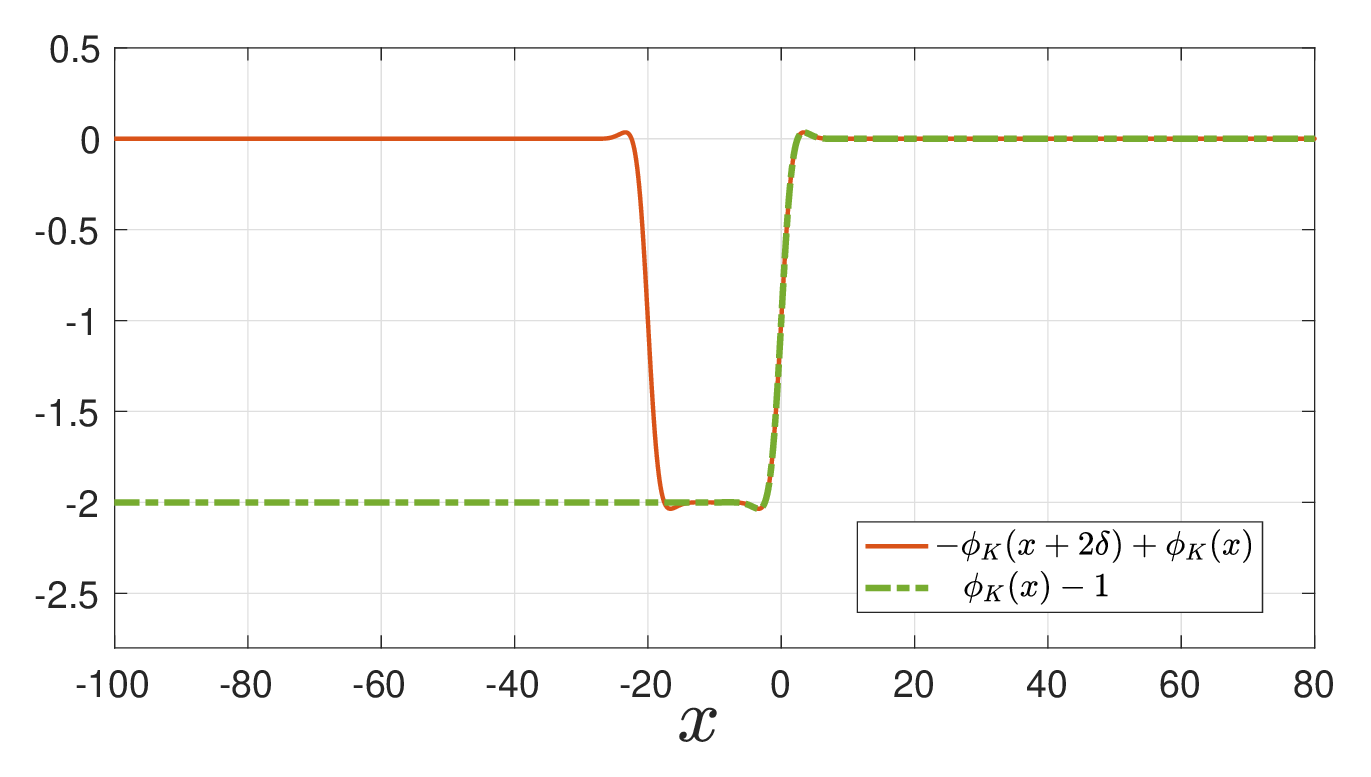}
\end{center}
\caption{Kink-antikink pair after being shifted left, moved up, and multiplied by $-1$ where $\delta = 10$ (solid red) and single kink translated down a unit (dashed green). Note the similarity of the right tails.}
\label{tailShift2}
\end{figure}

Next we consider the tail of a kink-antikink pair. Let us represent a single kink centered at $x=0$ by $\phi_K(x)$. We can then approximate a kink-antikink pair using 
\begin{equation}
\phi_K(x+\delta)-\phi_K(x-\delta)-1.
\label{kak}
\end{equation} 
Thus, $2\delta$ is the distance from the kink center to the antikink center. We then recenter the pair by translating leftward by $\delta$ units in the horizontal direction to put the position of the antikink at $x=0$, translate by $1$ in the vertical direction (upward) so that the tail lies on the $x$-axis, and finally multiply by $-1$ so that the tail has the same orientation as a single kink (making it easier to compare a single kink tail with a kink-antikink tail); the result is
\begin{equation}
-\phi_K(x+2\delta)+\phi_K(x).
\label{kak2}
\end{equation} 
See Figure (\ref{tailShift2}) for a pictorial representation of the
result along with a single kink moved down one unit given by $\phi_K(x)-1$.

Given that the tail of a single kink can be represented as ${\beta}/{x^\alpha}$ (for $x$ sufficiently large), the (right) tail of a kink-antikink pair (adjusted so that the tail lies along the $x$-axis) would be given by $\beta/(x+\delta)^\alpha-\beta/(x-\delta)^\alpha$. Shifting this to the left by $\delta$ units so that position of the antikink is at $x=0$, and multiplying by $-1$ (as we did with the entire kink-antikink pair) we get
\begin{equation}
\beta/x^\alpha-\beta/(x+2\delta)^\alpha
\label{kakTail}
\end{equation}
We can factor out the $\beta$ in Eq (\ref{kakTail}) and rewrite the expression as 
\begin{equation}
\beta\frac{(2\delta+x)^\alpha-x^\alpha}{x^\alpha (2\delta+x)^\alpha}
    \label{kakTail2}
\end{equation}
In Eq.~(\ref{kakTail}) the first term dominates for small $x$ and large $\delta$. For the expression in Eq.~(\ref{kakTail2}), as $x$ gets large for fixed $\delta$ the asymptotic form of the expression becomes  
\begin{equation}
\frac{2 \beta \delta \alpha}{x^{\alpha+1}}
    \label{kakTail3}
\end{equation}
which can be found by expanding the parenthetical expressions and keeping only the highest order terms. Thus, we expect that for small $x$ we would get a slope of $-\alpha$ and for large $x$ a slope of $-(\alpha+1)$. 

\begin{figure}[]
\begin{center}
\includegraphics[width=14cm]{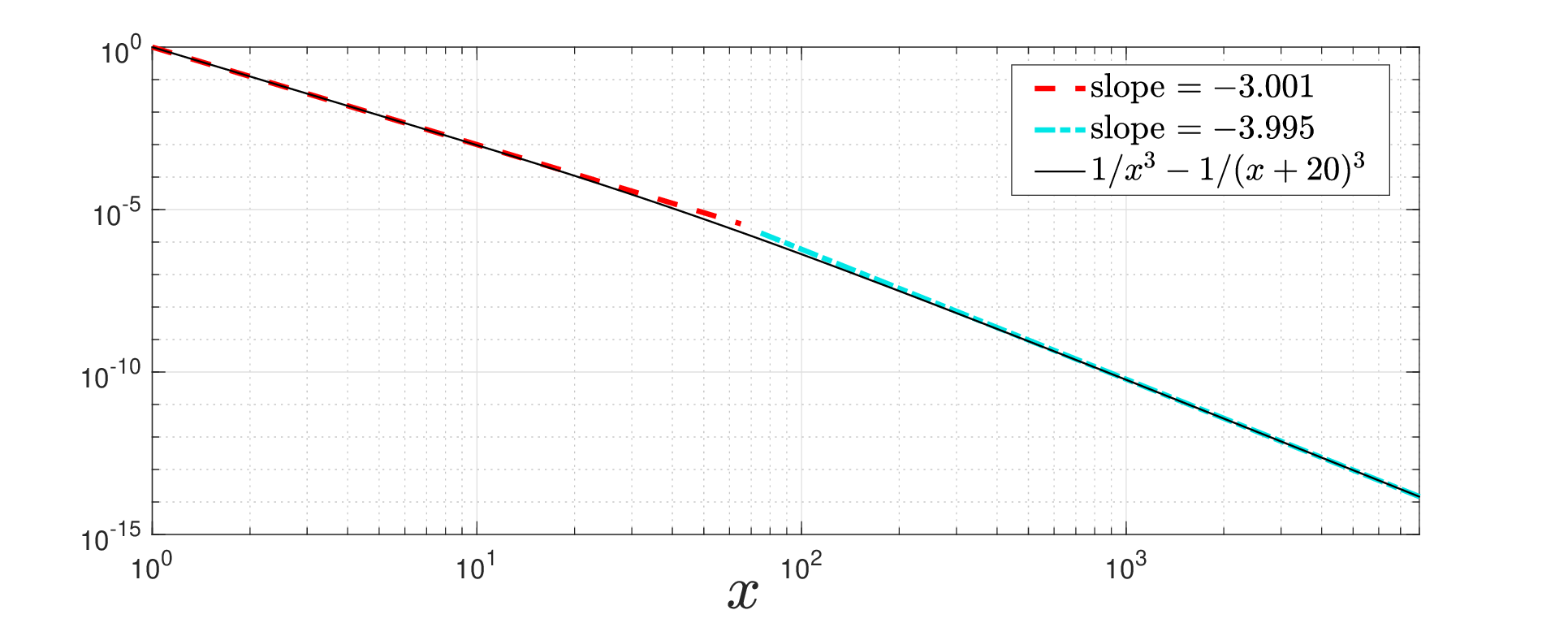}
\end{center}
\caption{Log-log plot of a ``tails only''  kink-antikink pair given by Eq. (\ref{kakTail}) with $\alpha=3$, $\beta=1$ and $\delta=10$ (in black). The red (dashed) line captures the slope ($-3.001$) of the first part of black line over approximately $1\leq x \leq 10$, and the cyan (dash-dotted) line captures the slope ($-3.995$) of the last part of the black line over approximately $ x \geq 1000$. Both red and cyan lines are regression best fit lines for the corresponding regions of the black line.}
\label{tailShift}
\end{figure}

We illustrate this in Figure~\ref{tailShift} with a log-log plot of the expression in Eq.~(\ref{kakTail}) (a ``tails only'' plot) 
using $\alpha=3$, $\beta=1$ and $\delta=10$ for $1\leq x \leq 8000$. It is clear that there is an initial nearly straight part, followed by a slight bend, and then another nearly straight part. We have highlighted the two straight parts of the curve. Using linear regression we find that the slope of the initial segment (red-dashed) is $-3.001$ and the slope of the final segment (cyan-dash-dotted) is $-3.995$. From Eq.~(\ref{kakTail}) we would have expected the slope of the first segment to be $-3$ and from 
Eq.~(\ref{kakTail3}) we would have expected the second slope to be $-4$, in very close agreement with the regression results.

\begin{figure}[]
\begin{center}
\includegraphics[width=8cm]{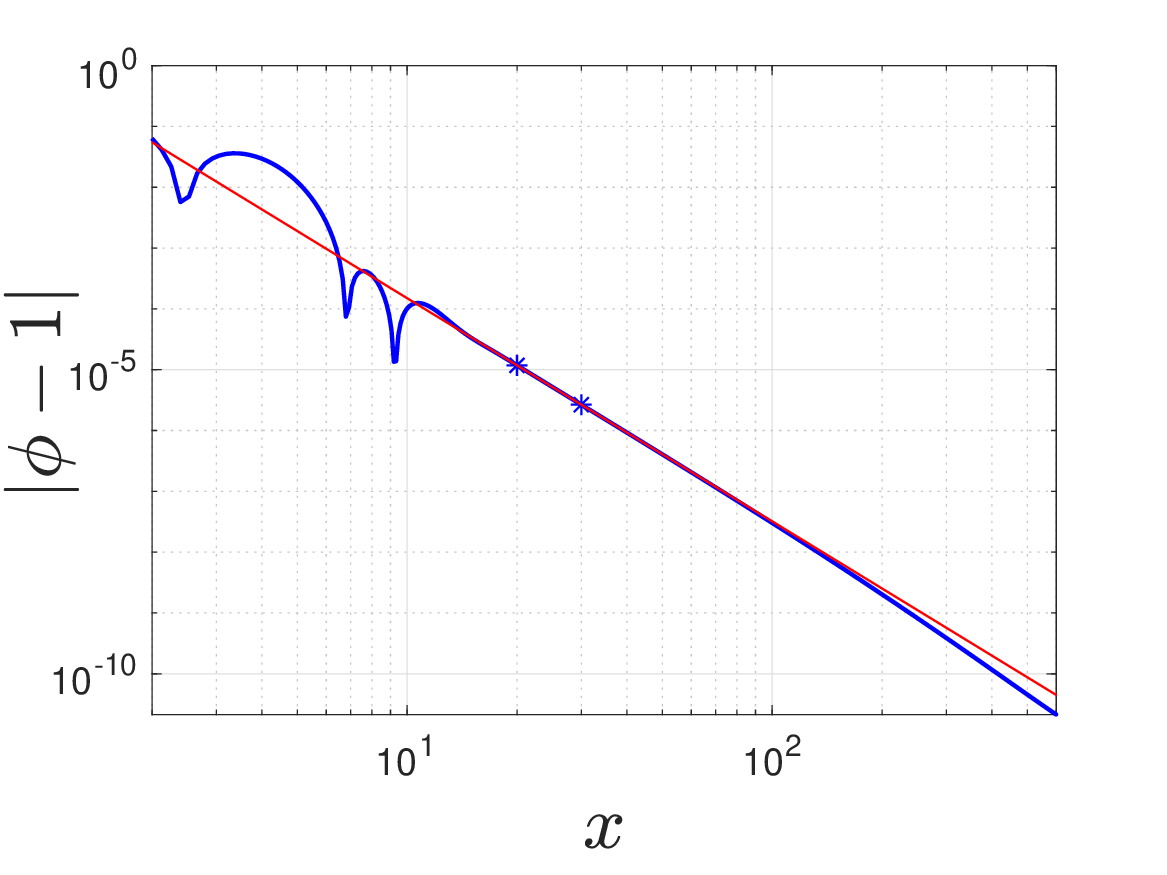}
\includegraphics[width=8cm]{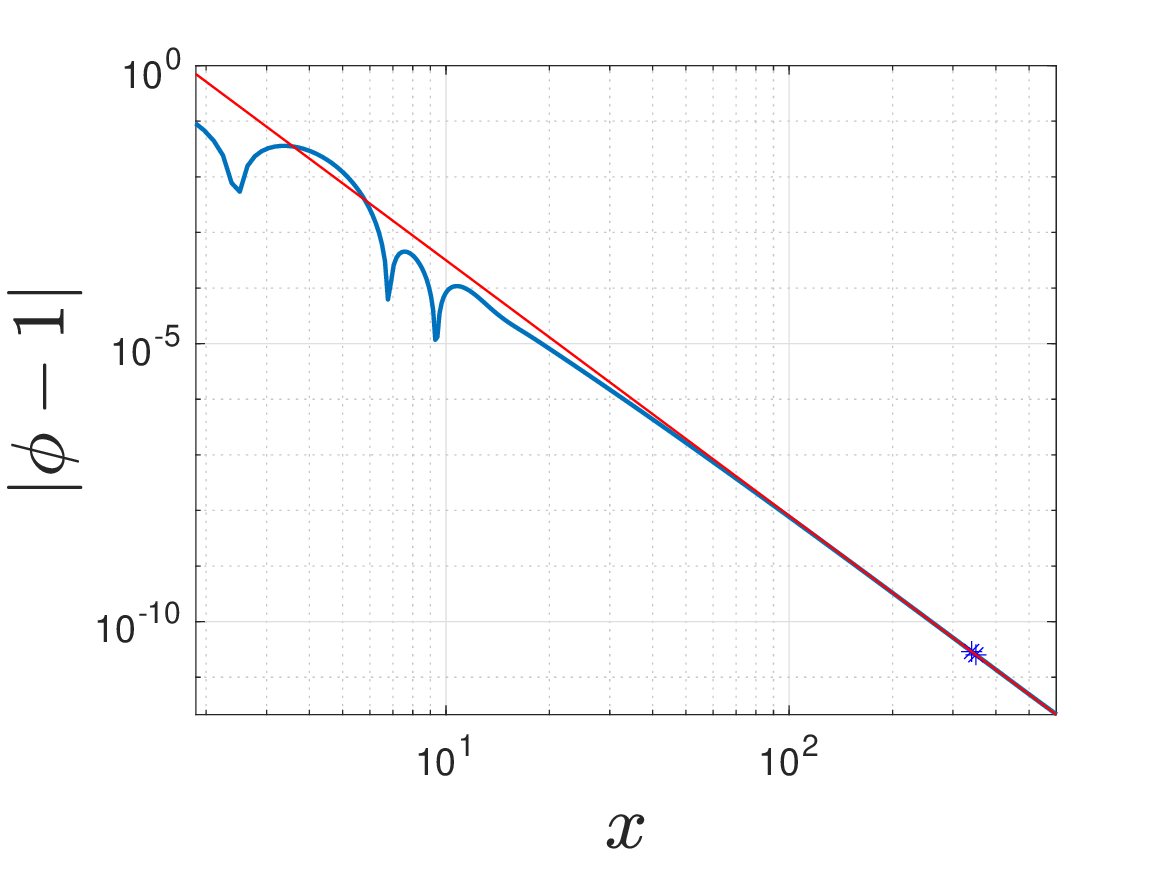}
\end{center}
\caption{Slope of tail of a kink-antikink pair, for $\alpha=3.65$ for two different kink-antikink separation values (in blue for both figures). Separation is $2\delta=100$ for left panel (which is a constrained state), and $2\delta=7.52$ for right panel (a steady state). The red lines are straight lines fit to the data between the blue asterisks (in the right panel the asterisks are hard to distinguish from each other due to the compression of the x-axis on a loglog plot). Slope of left panel red line is $-3.67$ (close to $\alpha$) and slope of right panel red line is $-4.60$ (close to $\alpha+1$). }
\label{KAKtail}
\end{figure}

We can observe the same effect in plots of actual kink-antikink tails, though it is less pronounced due to limitations in the length of the $x$-interval that can practically be used, the fact that the tail approximation does not start immediately at $x=0$, and the effects of finite boundary conditions used to simulate infinite boundaries. Thus we require two separate figures to show the same effect as we see in Figure \ref{tailShift}. In Figure \ref{KAKtail} we show the right tail of two kink-antikink pairs using log-log plots. Both correspond to $\alpha=3.65$, but for the first (left panel) the distance between kink and antikink is $2\delta=100$ and for the second (right panel) the distance is $2\delta=7.52$. 

We expect that in the left panel the large separation dominates and we should get a value of around $-3.65$ for the slope immediately following the oscillation phase. In this case the first term of Equation (\ref{kakTail}) dominates up to approximately $x=100$. The calculated value of the slope over the region between the asterisks is $-3.67$, close to the theoretical value of $\alpha=-3.65$. Beyond $x=100$ for this case the slope has just begun its dip towards the long-term value of $\alpha+1=-4.65$. For the right panel, since the separation is quite small, we should enter the asymptotic region ($x \rightarrow \infty$) of the tail fairly quickly. Indeed the value of the slope over the region between the asterisks (which are hard to distinguish from each other due to distortion of the $x$ scale in a log-log plot) is $-4.60$, close to the predicted value of $\alpha=1=4.65$ as Eq.~(\ref{kakTail3}) now comes into play for large $x$. 

Notice that for kink-antikink tails, there can be three regions, depending on the values of $\alpha$ and the separation parameter $\delta$. We can refer to these as the ``oscillation region'', the ``$\alpha$ region'' and the ``$\alpha+1$ region''. The $\alpha+1$ region will always exist for large $x$, but may be difficult to observe numerically as in the left panel of Figure \ref{KAKtail}. When oscillations exist (for $\alpha>3.567$ or so), the $\alpha$ region can be ``crowded out'' by the oscillation region and hence not be readily observable as in the right panel of Figure \ref{KAKtail}. In this setting of 
$\alpha>3.567$, the innermost 
region always demonstrates
the relevant oscillations.

\section{Kink-antikink interactions - PDE and ODE simulations}
\label{sec:accel}
For a kink-antikink pair, the half separation distance is the same as the position of the antikink, which we denote $x_{AK}(t)$. For a given initial half separation $\delta=x_{AK}(0)$, we can calculate the initial acceleration $\ddot x_{AK}(0)$ (proxy for the
force between the two structures) as described in Section \ref{zeroVelocityStates}. Specifically, for a given $\delta$, let $\phi_0(x)$ represent the result of the constrained minimization described in Section \ref{zeroVelocityStates}, using Equation (\ref{KAKinitial}) (along with the given $\delta$) as the initializer. Then we can use $u(x,t=0)=\phi_0(x)$ and $u_t(x,t=0)=0$ as initial conditions for a numerical simulation of Equation (\ref{beam1}), over a short interval of time ($0\leq t \leq 0.01)$. The antikink position $x_{AK}(t)$ is given by the positive $x$ value of the intersection of $u(x,t)$ with the $x$-axis. We can then calculate (using numerical differentiation) the antikink velocity $\dot x_{AK}(t)$ over $0\leq t \leq 0.01$, and the slope of the $\dot x_{AK}(t)$ graph will give the initial acceleration $\ddot x_{AK}(0)$. Now that we have the acceleration as a function of $\delta$, which we can denote $a(\delta)$, we can numerically integrate $-a(\delta)$ to get potential energy as a function of $\delta$, which we denote $U(\delta)$.

\begin{figure}[]
\begin{center}
\includegraphics[width=8cm]{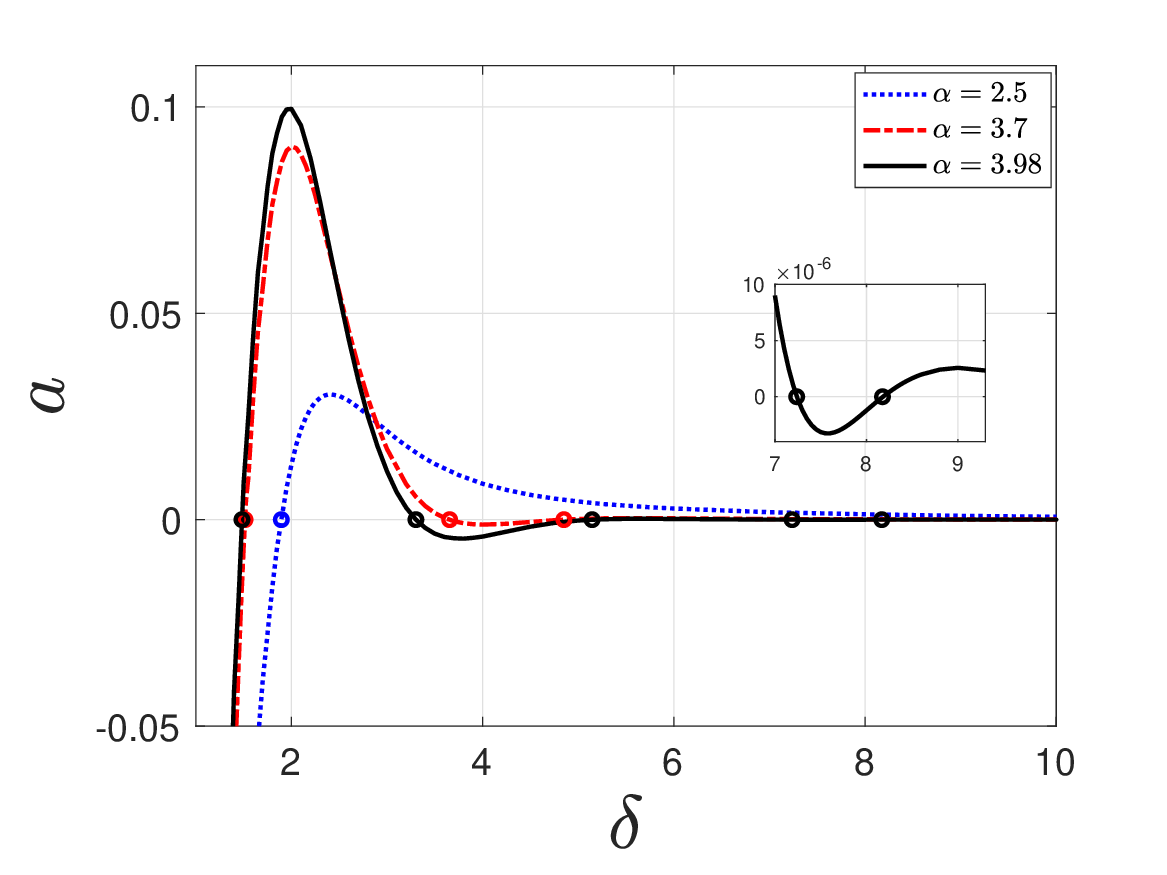}
\includegraphics[width=8cm]{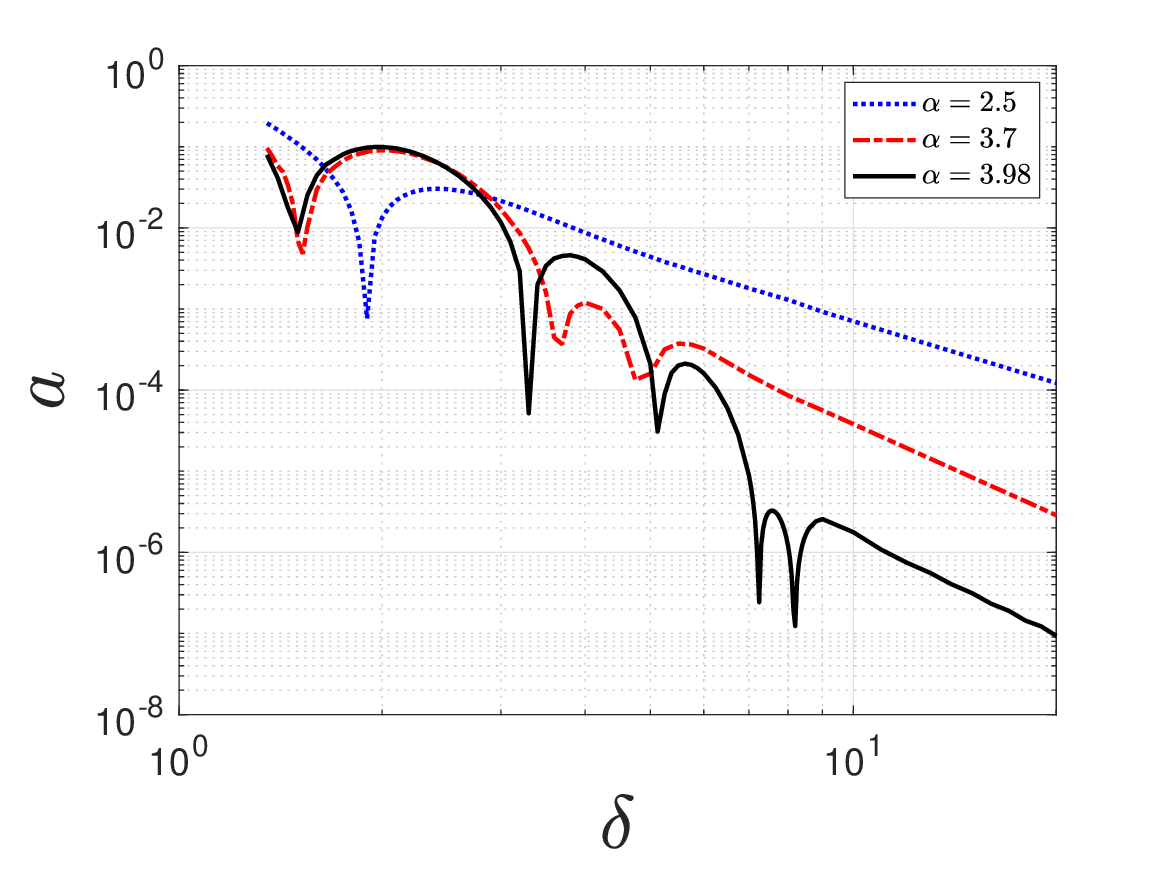}
\includegraphics[width=8cm]{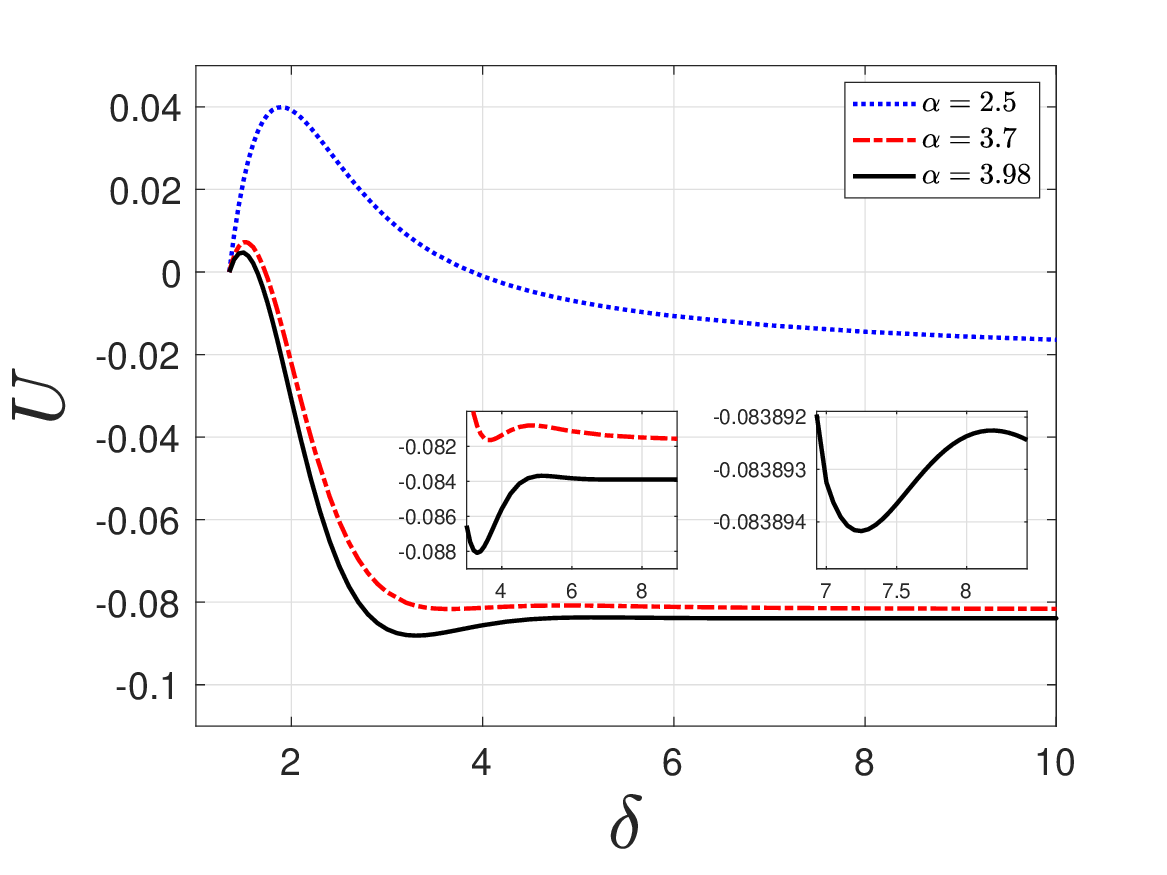}
\end{center}
\caption{Three views of acceleration and its negative antiderivative (potential energy landscape) as a function of antikink position (half-separation distance) for $\alpha=2.5$ (blue-dotted), $\alpha=3.7$ (red-dash-dotted), $\alpha=3.98$ (black-solid). Top left: antikink acceleration ($a$) vs antikink position with zeros of curves marked with circles. Top right: loglog plot of the absolute values of the curves in the top left. Bottom: negative of the antiderivative of the top left (potential energy, $U$). Note that in the top right panel, we can discern an oscillatory region (modulated by a decaying exponential) for small separations followed by a power law decay region.}
\label{accel-energy}
\end{figure}

This was done for three values of derivative order $\alpha$=2.5, 3.7 and 3.98. The data is shown in various forms in Figure \ref{accel-energy}. We see that the form of the acceleration graphs (top two panels) closely resembles the graphs of the single kink tails shown in Figure \ref{tails2}.
The bottom figure, importantly, presents the potential
energy landscape (negative of the integral of the acceleration) that one (anti)kink experiences in the presence
of the kink. This landscape is responsible for the phase-portrait
of the kink-antikink interaction dynamics as will be shown below
through comparison of the PDE with an effective derived 
one-degree-of-freedom ODE.

We can model this data as we did with kink tails, using an exponentially modulated oscillation for small separation values, and a power function for the larger separation values. For the given $\alpha$ values we get power law models for the acceleration of the form $a(\delta)=b/\delta^k$ where $k$=2.53, 3.72, and 4.00 respectively. This is clearly suggestive of the possibility that $k=\alpha$, similar to the case of the kink tail, though the numerical evidence is not quite as definitive.

\begin{figure}[]
\begin{center}
\includegraphics[width=8cm]{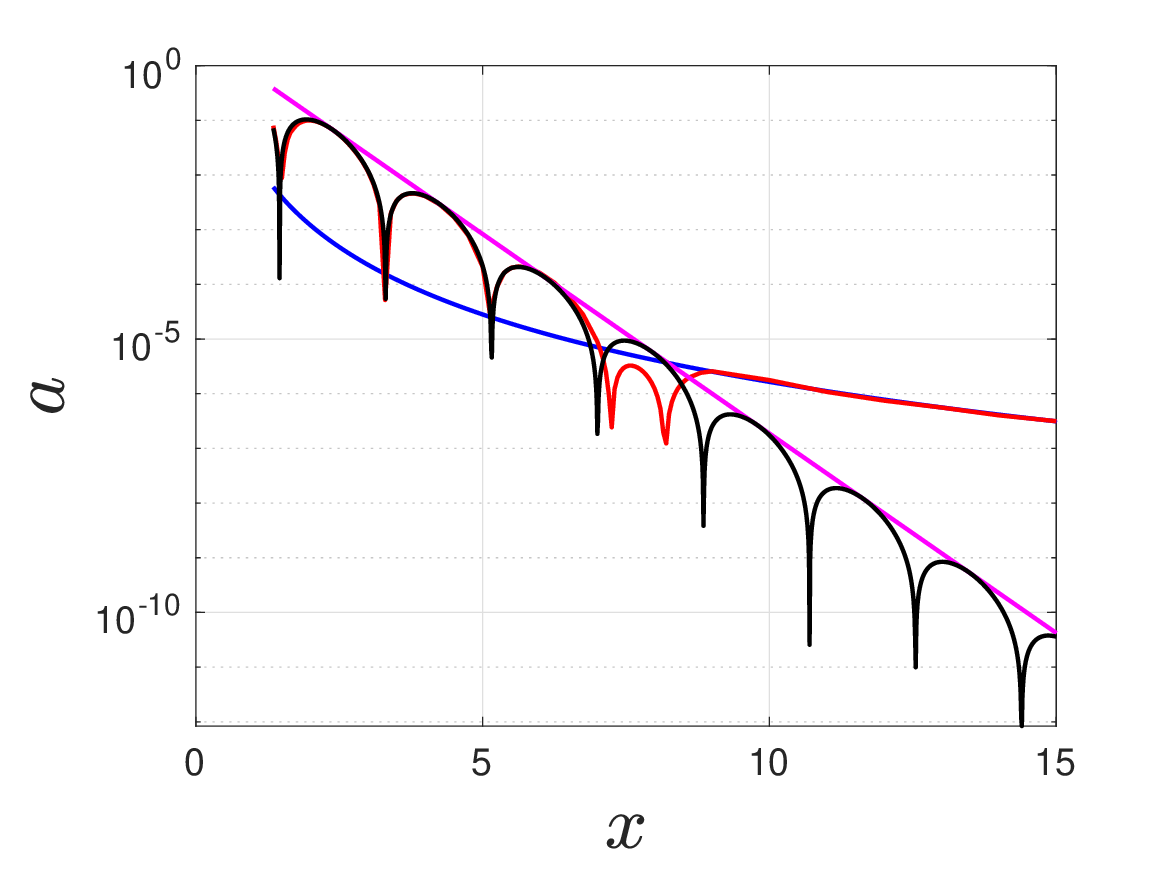}
\includegraphics[width=8cm]{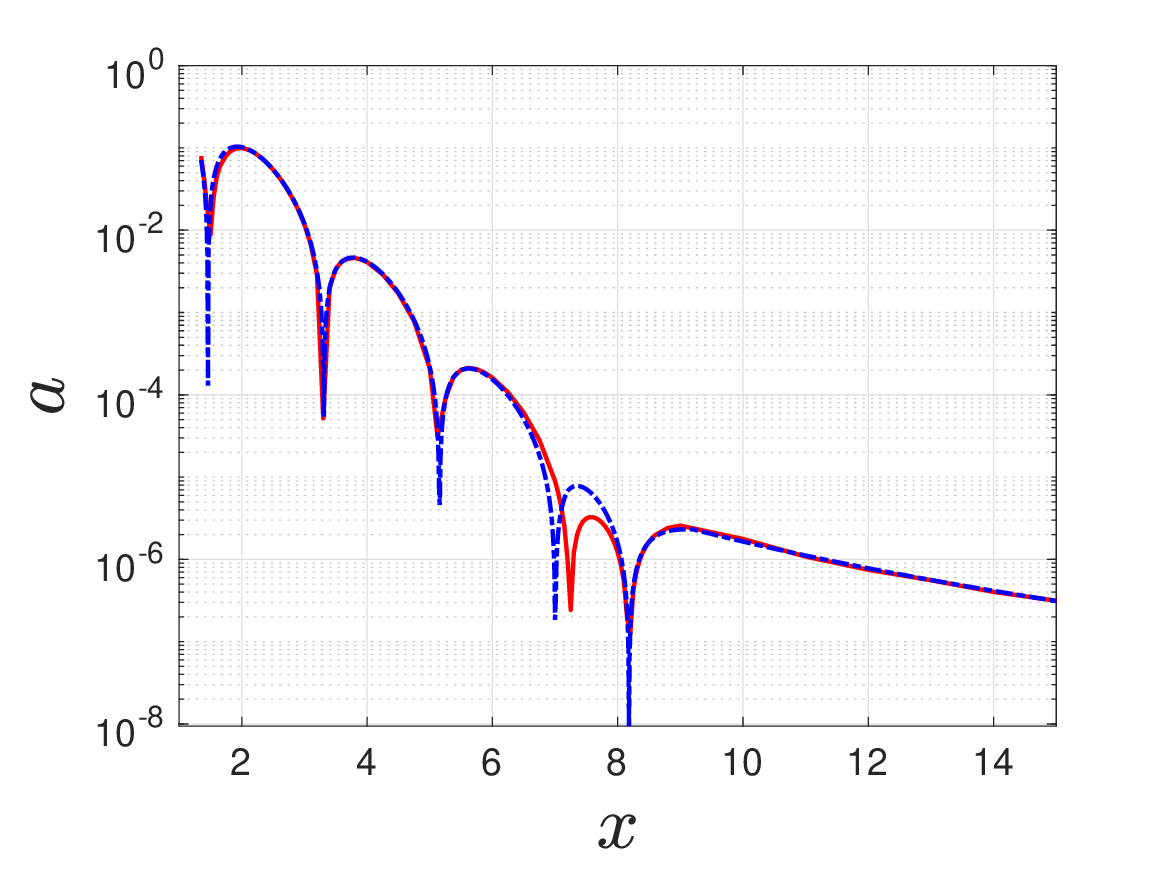}
\end{center}
\caption{Modeling the acceleration of antikink position and a consequent ODE model for alpha=3.98. Note that we are using $x$ in place of $x_{AK}$ (or $\delta$) for the kink-antikink half separation distance. As we saw above for the tails when alpha=3.99 (see Figure \ref{tails2}) we can develop models for the oscillatory part of the acceleration (small values) and for the non-oscillating part (larger acceleration values). In the left panel we can identify a point where the power model ($y=0.0206/x^{4.097}$ in blue) intersects the envelope curve of the oscillation ($y=-3.7\exp(-1.68x)\sin(1.7x+.66)$
in black with envelope in magenta), as occurring at a separation value of $x=8.2$. Based on this we created a model $M(x)$ which ``stitches together'' the power and exponential/oscillatory model, and which is shown in the second figure (in blue) with the data shown in red. This can then be used to create an (approximate) ODE model for the half separation distance given by Eq. (\ref{odeMOdel}).}.
\label{accelModel}
\end{figure}

For the case $\alpha$=3.98 we show the results of fitting both parts (exponentially decaying oscillation and power law decay) in Figure \ref{accelModel}. In that figure we show the two models separately (left panel) and also a single model that ``stitches'' together the two parts (right panel). The stitched model uses two linear functions on a transition interval $[c,d]$ 
which contains the point where the envelope curve of the oscillation meets the power law decay curve. It is convenient and consistent with standard function notation at this point to drop the subscript on the antikink position and use $x$ in place of $x_{AK}$ (or $\delta$) as the independent variable (i.e. the half separation distance). The first such linear function, $\phi_1(x)$, goes from 1 to 0 on the interval $[c,d]$, and the second, $\phi_2(x)$, goes from 0 to 1 on the interval. The fitted model for the oscillation part is given by $f_1(x)=-3.7\exp(-1.68x)\sin(1.7x+.66)$ and the fitted model for the power law part is given by $f_2(x)=0.0206/x^{4.097}$. Then the expression $M(x)=f_1(x) \phi_1(x)+f_2(x) \phi_2(x)$ approximates the entire data set, with the transition from oscillation to power occurring between $c$ and $d$. 
As can be seen in the right panel of the figure,
this approximation is fairly accurate throughout
$x>0$, with the largest deviations occurring in a
small interval near the transition region.

Using the stitched model $M(x)$ for the acceleration we can create an ODE that mimics the behaviour of the PDE, using

\begin{equation}
\ddot{x}=M(x). 
\label{odeMOdel}
\end{equation}

\begin{figure}[]
\begin{center}
\includegraphics[width=5.5cm]{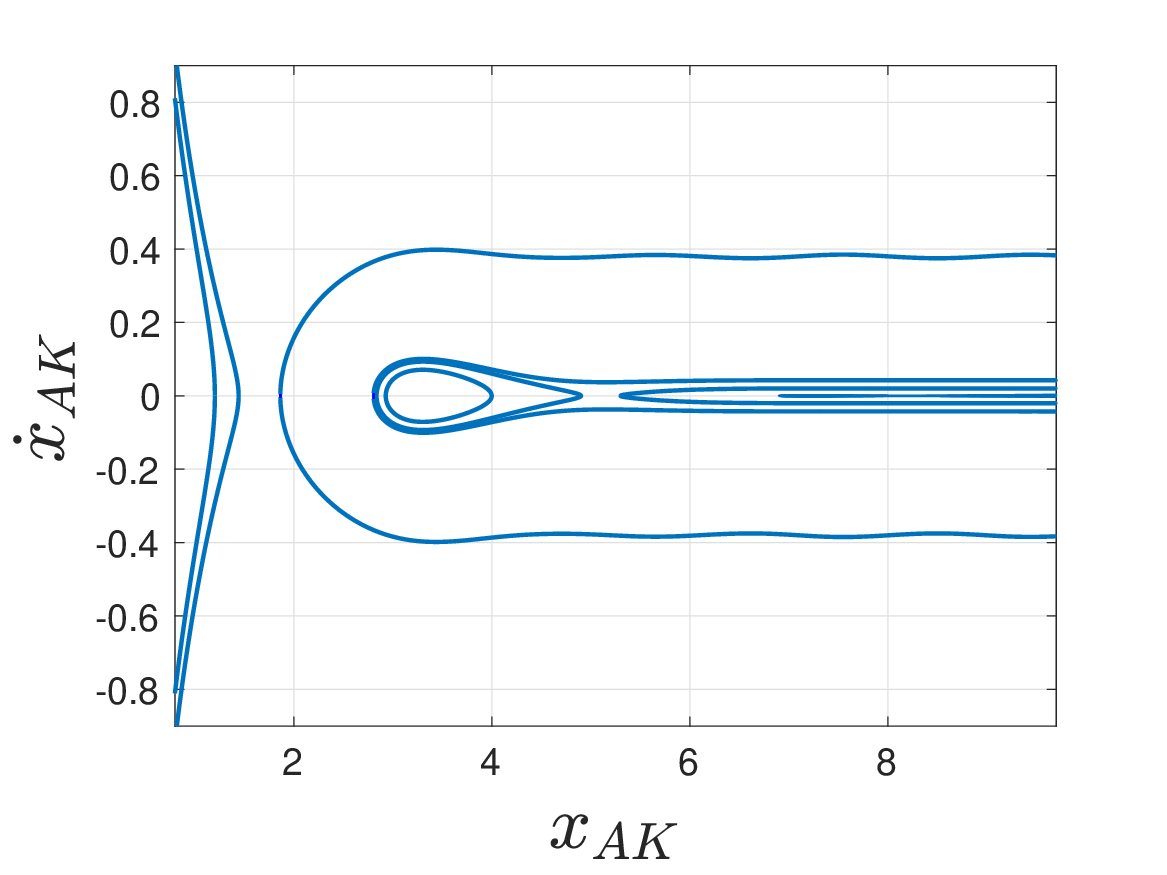}
\includegraphics[width=5.5cm]{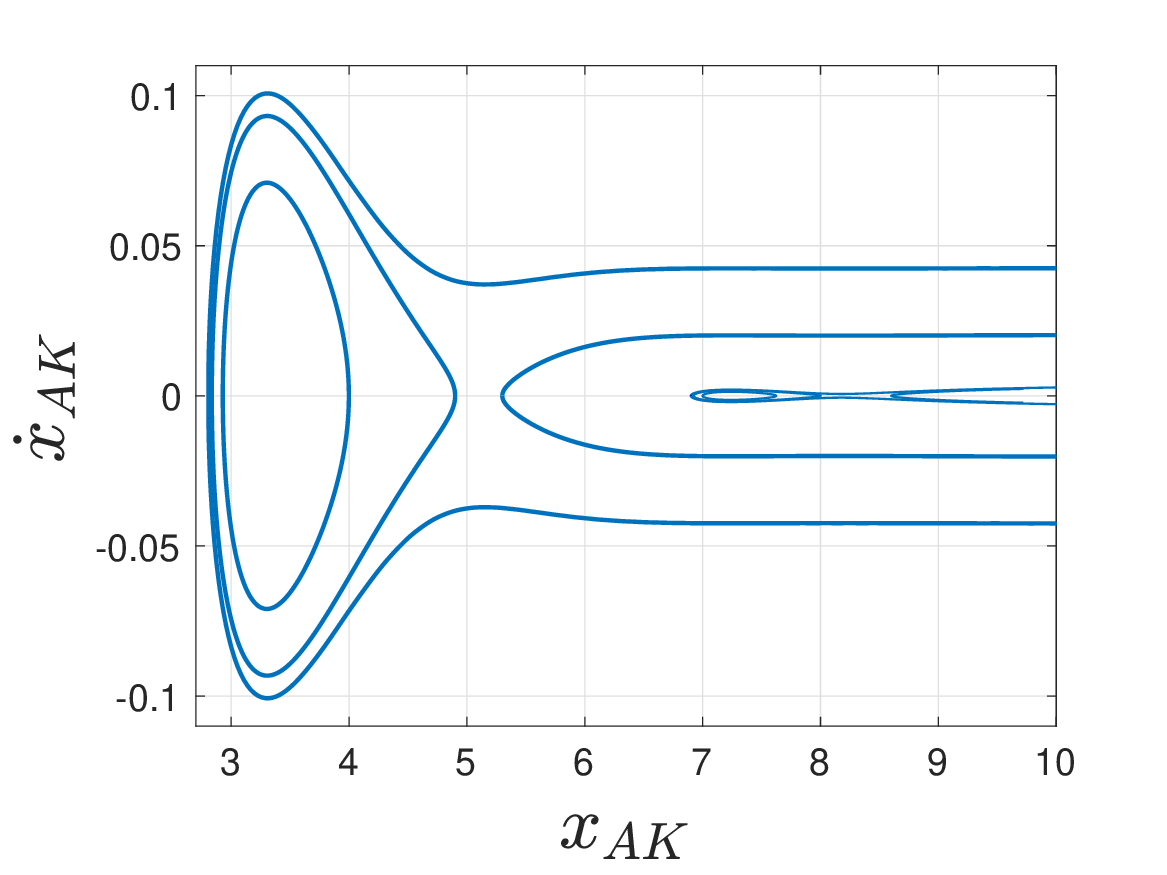}
\includegraphics[width=5.5cm]{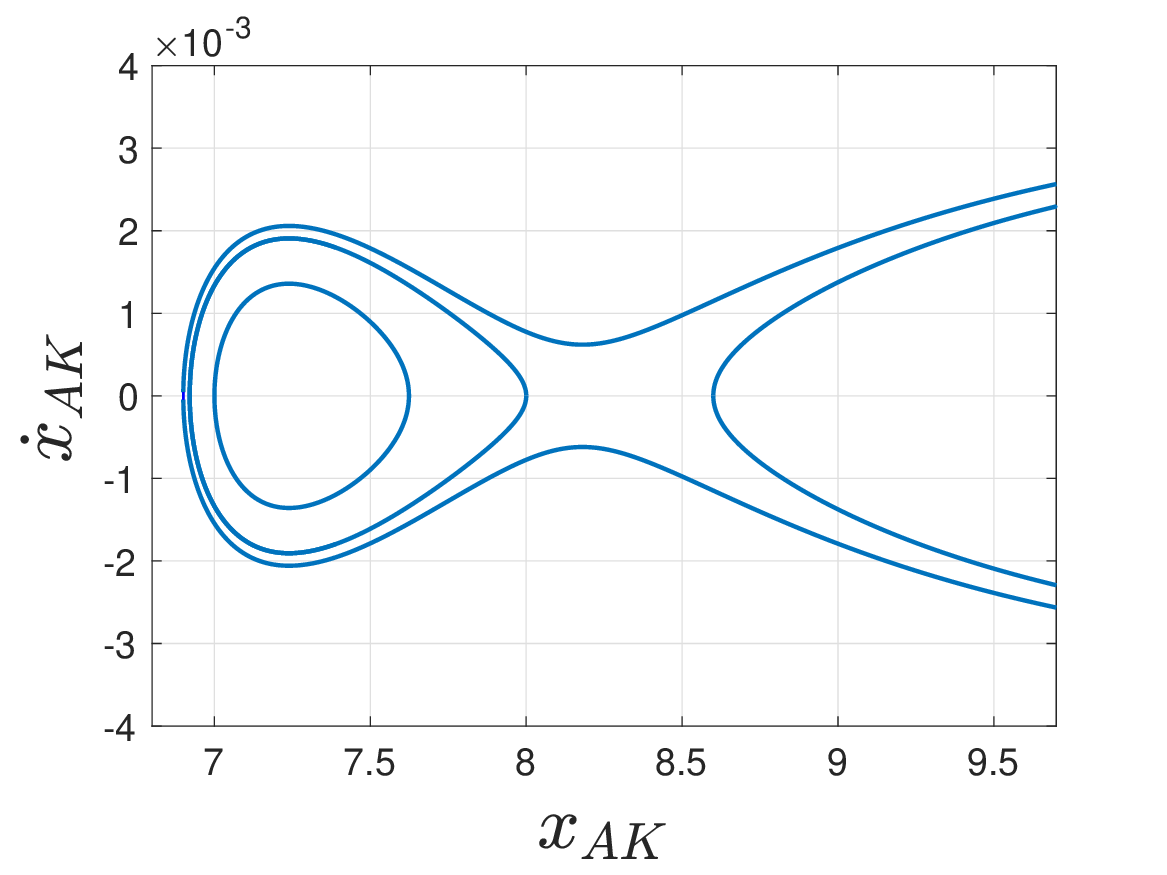}
\includegraphics[width=5.5cm]{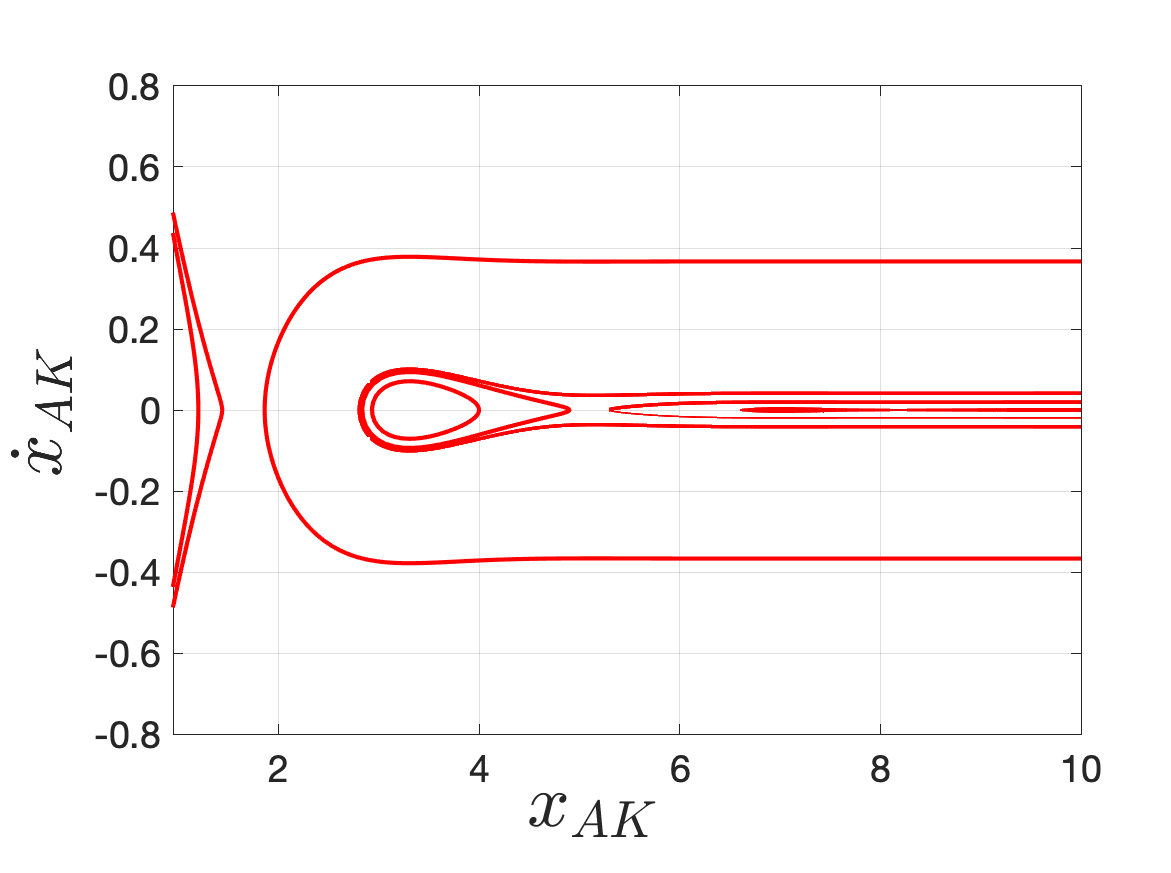}
\includegraphics[width=5.5cm]{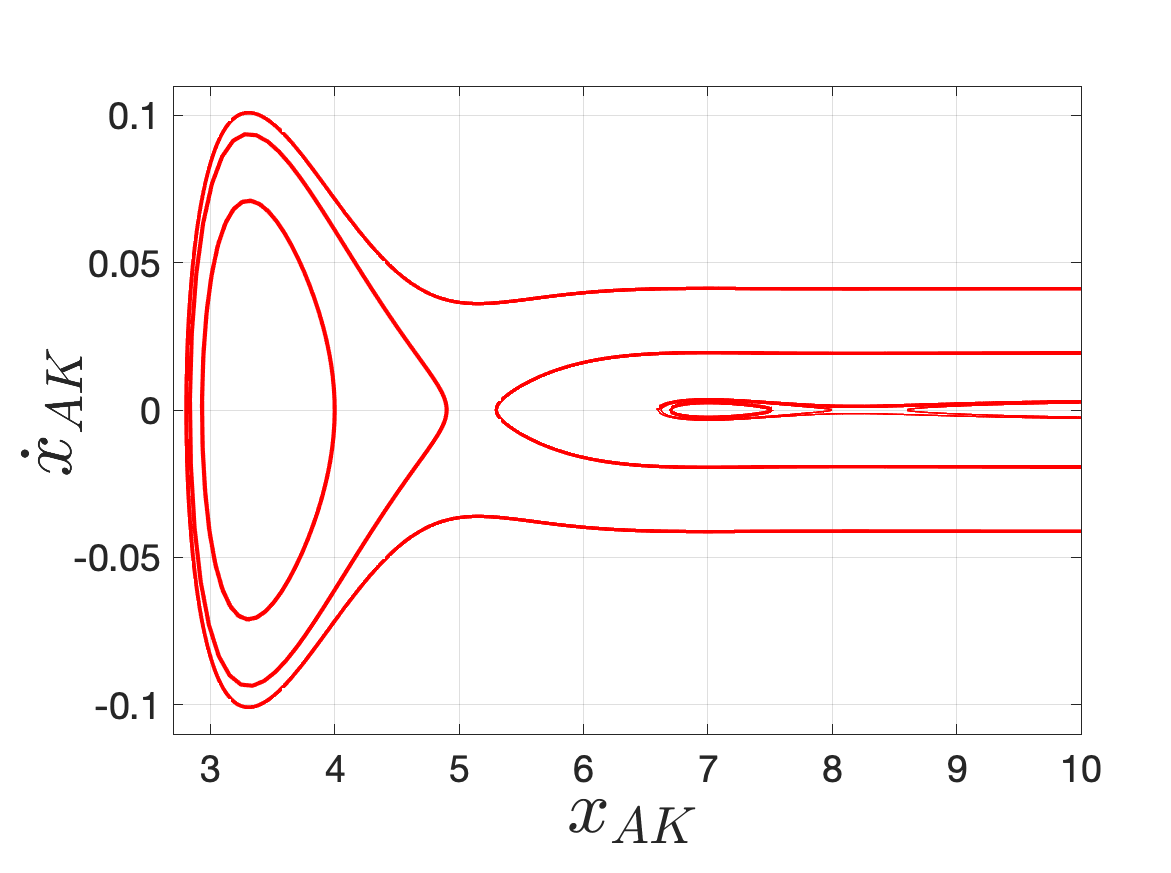} \includegraphics[width=5.5cm]{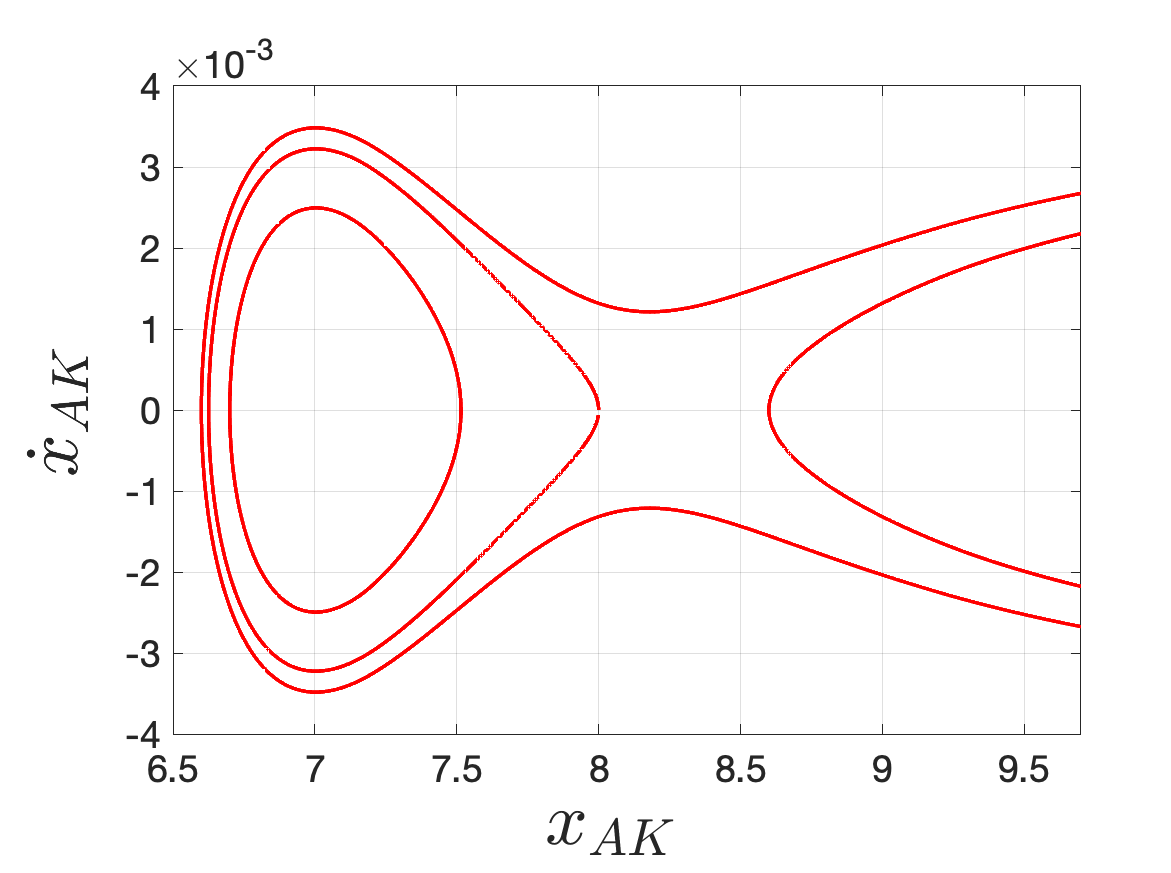}
\end{center}

\caption{Dynamics of the PDE (top row) and corresponding ODE (bottom row), for $\alpha=3.98$, expressed as phase portraits with successive zooms going left to right. For the PDE model $x_{AK}(t)$ is calculated as the positive $x$ intersection of the solution to Eq. (\ref{beam1}) with the horizontal axis, and the ODE model is the solution to Eq. (\ref{odeMOdel}). In all cases we are plotting the antikink position $x_{AK}(t)$ on the horizontal axis and $\dot x_{AK}(t)$ on the vertical axis. The PDE and ODE models give similar results, though the ODE model is somewhat inaccurate near the separation value of 8.2 where the ``stitching" of the two models has been performed (apparent in the difference between the 
right panels of the two rows).}
\label{phaseP}
\end{figure}

For the current case of $\alpha=3.98$, the envelope curve for the oscillation and the power curve 
have an intersection at the value  $x=8.2$, and we use $c=7.2$ and $d=9.2$ for the transition interval $[c,d]$. Dynamics for both the PDE and the ODE model are shown in Figure \ref{phaseP} in the form of phase portraits
(associated also with the potential energy
landscape of Fig.~\ref{accel-energy}).
A very good agreement is identified between the two suggesting
that the stitched model is an adequate 
and interpretable (i.e., encompassing the
main ingredients of kink-antikink interactions) description of the kink-antikink
dynamics in all of space. Naturally, there are
some quantitative deviations between the two. E.g., note
the specifics of the right panel phase portraits,
especially around the left center point in Fig.~\ref{phaseP}, which reflects the above
mentioned transition region. Yet, there is an excellent 
qualitative and overall good quantitative 
agreement.

\section{$\alpha$ values outside the range $2 \le \alpha \le 4$}

\begin{figure}[]
\begin{center}
\includegraphics[width=8.5cm, height=6cm] 
{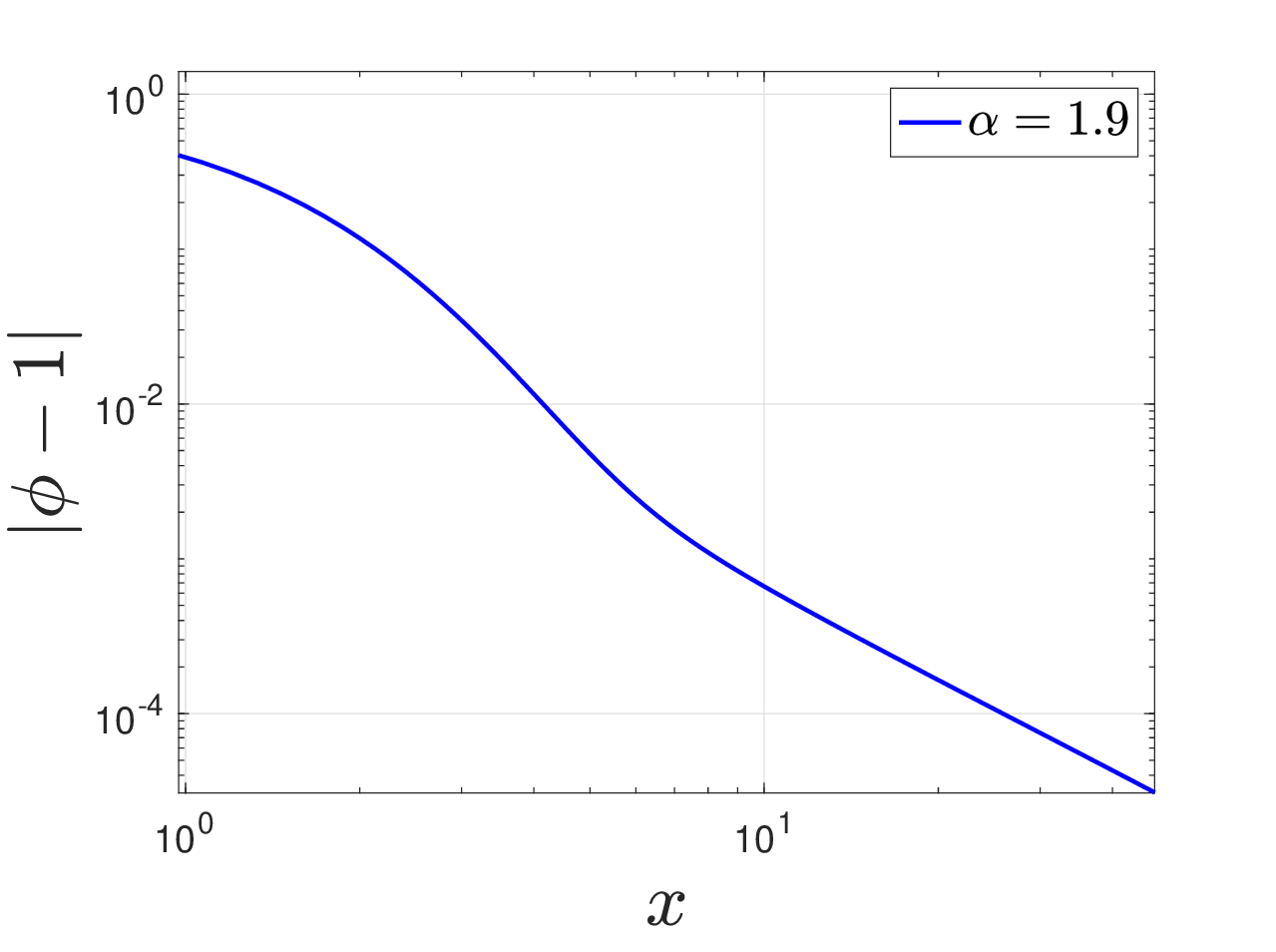}
\includegraphics[width=8.5cm, height=6cm]{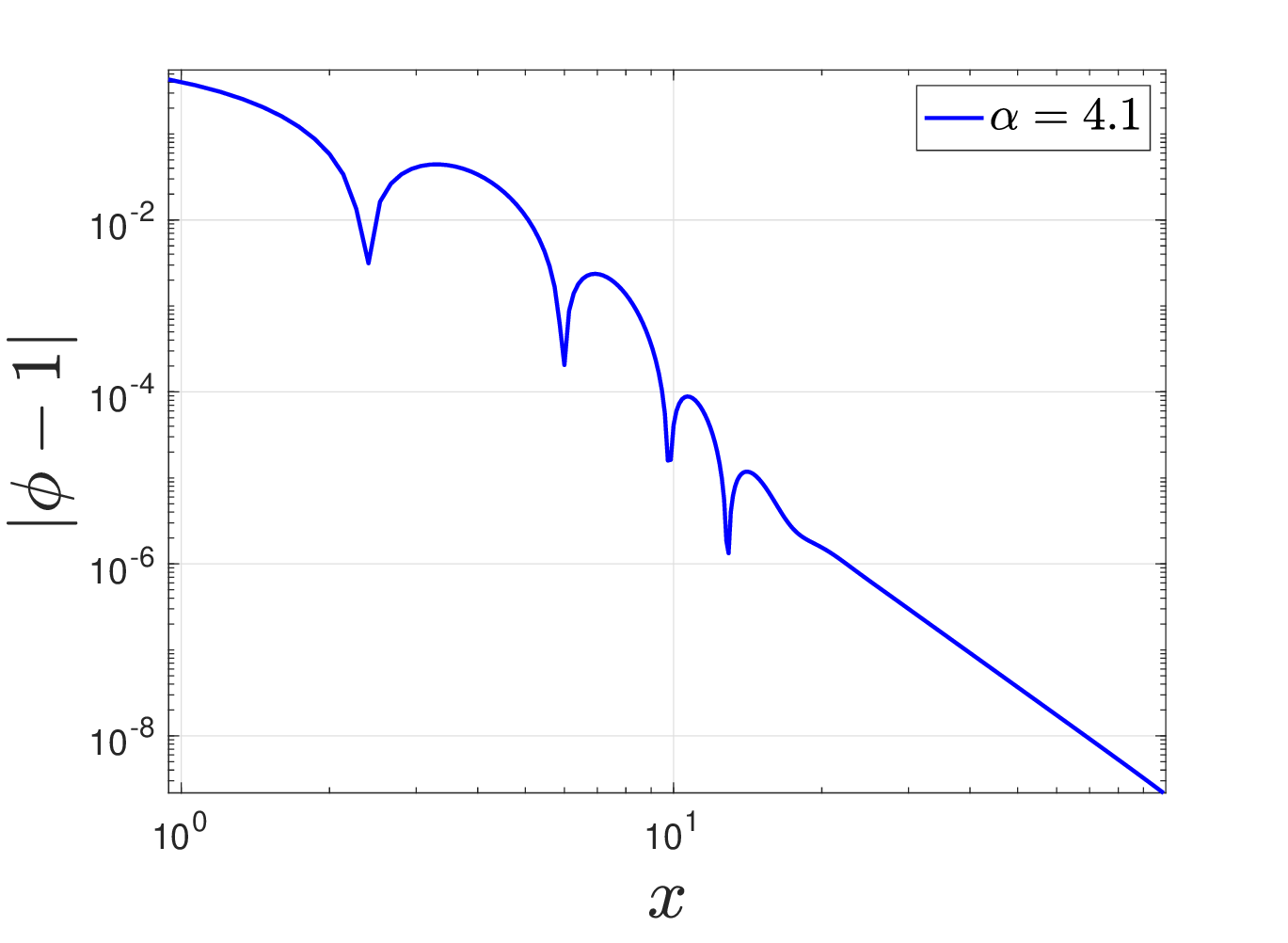}
\end{center}
\caption{Log-log plots of single kinks for $\alpha$ slightly less than 2 or greater than 4. Left is $\alpha=1.9$ and right is $\alpha=4.1$}
\label{otherAlphaKinks}
\end{figure}

Finally, we have also explored the kink dynamics outside
of the continuation region of $\alpha \in [2,4]$ discussed above;
see Figure \ref{otherAlphaKinks} for log-log plots of single kinks (shifted down one unit) for $\alpha=1.9$ and $\alpha=4.1$. For $\alpha$ slightly less than 2, we see that the kink has a power law tail (as for kinks with $\alpha$ between 2 and 4, but unlike for $\alpha=2$ which has an exponential tail). 
However, it is important to note that there are no crossings of the y-asymptote at $y=1$, as there exist for $\alpha$ between 2 and 4. For $\alpha$ slightly greater than 4, the shape of the kink resembles that of a kink when $\alpha$ is close to and slightly less than 4, with a power-law tail and several crossings of the asymptote. Thus,
the infinite number of crossings of the case with $\alpha=4$ is
lost, reverting back to a finite number of crossings prior to
a power-law tail. 

As a result of the tail shapes, we expect that there are no kink-antikink steady states for $\alpha$ values slightly smaller than 2, but that the steady states for $\alpha$ slightly larger than 4 the bifurcation diagram in Figure \ref{bifurcate1} continues in a similar manner as for $\alpha$ slightly smaller than 4.

\section{Conclusions and Future Challenges}

In the present work we considered
the effect of a spatial Riesz fractional
derivative in a real, field-theoretic
$\phi^4$ model. This enabled us to
{\it interpolate} between the harmonic
and the biharmonic variants of the model.
The use of the Riesz spatial derivative
allows us to consider the relevant boundary
value problem in a continuous way, interpolating
between the two limits and appreciating how
one morphs into the other upon continuation
of the relevant exponent $\alpha$.
Indeed, what this shows is that while 
the biharmonic limit features infinitely
many kink-antikink pairs, the vast majority
of them disappears exponentially close to 
the biharmonic limit through a sequence
of saddle-center bifurcations. Only one such
pair ``survives'' for $\alpha<3.96$, eventually
disappearing around $\alpha \approx 3.57$.
Thereafter, only a single branch of
kink-antikink pairs survives, namely the one associated
with the smallest equilibrium distance
between the two structures. Remarkably, such
an isolated equilibrium state cannot disappear
(bifurcation-wise) by itself and hence has the
kink-antikink separation diverges to infinity
as the harmonic limit is approached, i.e., one
can say that this branch emerges bifurcating
from infinity at the harmonic limit.
We have also explored both the tails
of individual kinks (finding a direct 
correspondence thereof with the fractional
exponent) and those of kink-antikink
pair states, for which we observed and
argued that they decay asymptotically as a power law with
exponent $\alpha+1$. Such states also feature additional
regions, including a potential innermost oscillatory
one (depending on the value of $\alpha$) and
one of a power law 
with exponent $\alpha$. Additional features,
including the kink-antikink interaction
force, as well as the dynamics for values
of
$\alpha$ outside the harmonic-biharmonic
range of $[2,4]$ were also considered.

Importantly, this work paves the way for a wide
range of future investigations. On the one hand,
from an analysis perspective, many of the features
established numerically herein would be particularly
worthwhile to seek to quantify in analytical detail. 
For instance, establishing the behavior near the harmonic
limit both above (single asymptote crossing) and below
(pure power law) would be of interest. Similarly,
establishing the dual (inner oscillatory exponential
and outer power law) behavior near the biharmonic
limit would also be of interest. Estimating the 
critical points of the relevant saddle-center bifurcations,
or developing more accurate or theoretically derived 
ODE models of kink-antikink interactions are all points
worthwhile of further investigation.

Our expectation is that this investigation
may pave the way for numerous additional
studies in the same spirit in additional
(e.g., other nonlinear Klein-Gordon ones), more complex
(and potentially higher-dimensional) models.
A prototypical example thereof is
naturally the nonlinear Schr{\"o}dinger
model, among others~\cite{eilbeck,dauxois}.
A different but equally promising direction
is that of tailoring dispersion in 
experimentally relevant settings~\cite{BlancoRedondoNC2016,RungeNP2020}
in order to achieve dispersion relations featuring non-integer dispersion. Ongoing 
discussions with experimental groups suggest
that this ground-breaking step may be within
reach. If this materializes, it promises to
be a stepping stone for extensive
(multi-component, multi-dimensional and other)
theoretical investigations of fractional
dispersive wave models~\cite{cuevas}. 

\appendix\section{Fractional Derivative Definitions}
\label{app:fracDef}
Fractional derivatives are nonlocal by nature; they
are defined over a given interval, not at a given point, as they are generally
based on integrals. For a given interval, each has a \textquotedblleft
Left\textquotedblright\ and \textquotedblleft Right\textquotedblright\
version. These correspond to either integrals with the lower limit of
integration fixed ($a$) and the upper limit the variable $(x)$ (Left
derivative), or with the lower limit of the integral variable $(x)$ and the upper limit fixed 
$(b)$ (Right derivative). 
We will use $D_{L}^{\alpha }$ and $%
D_{R}^{\alpha }$ to denote the left and right derivatives of order $\alpha $. For the Riemann-Liouville and Caputo derivatives, $n$ represents $\alpha $
rounded up to the nearest integer, so that $n-1<\alpha \leq n$. The definitions are as follows: \newline

\emph{Riemann-Liouville }

Left derivative: $(D_{L}^{\alpha }f)(x)=\frac{1}{\Gamma (n-\alpha )}\frac{%
d^{n}}{dx^{n}}\int_{a}^{x}(x-\tau )^{n-\alpha -1}f(\tau )d\tau $

Right derivative: $(D_{R}^{\alpha }f)(x)=\frac{(-1)^{n}}{\Gamma (n-\alpha )}%
\frac{d^{n}}{dx^{n}}\int_{x}^{b}(x-\tau )^{n-\alpha -1}f(\tau )d\tau $
\newline

\emph{Caputo}

Left derivative: $(D_{L}^{\alpha }f)(x)=\frac{1}{\Gamma (n-\alpha )}%
\int_{a}^{x}(x-\tau )^{n-\alpha -1}f^{(n)}(\tau )d\tau $

Right derivative: $(D_{R}^{\alpha }f)(x)=\frac{(-1)^{n}}{\Gamma (n-\alpha )}%
\int_{x}^{b}(x-\tau )^{n-\alpha -1}f^{(n)}(\tau )d\tau $\newline

\emph{Fourier}

Left derivative: $(D_{L}^{\alpha }f)(x)=F^{-1}[(ik)^{\alpha }\widehat{f}(k)]$

Right derivative: $(D_{R}^{\alpha }f)(x)=F^{-1}[(-ik)^{\alpha }\widehat{f}%
(k)]$\newline

where $\widehat{f}(k)=\int_{-\infty }^{\infty }\exp (-ikx)f(x)dx$ is the
Fourier transform of $f(x)$ and $F^{-1}$ denotes the inverse Fourier transform. \vspace{0pt}

It can be shown that as $a\rightarrow -\infty $ and $b\rightarrow \infty $, all
three definitions agree~\cite{podlubny1999fractional,exactResults2018} thus for our simulations of an infinite domain we only need the Fourier version (which is also the
fastest to implement numerically). Though the definitions for the left and right derivatives for the Fourier
case do not include \textquotedblleft left\textquotedblright\ or
\textquotedblleft right\textquotedblright\ integral, we will still use this
terminology since they agree with the left and right integral definitions
for the Riemann-Liouville and Caputo cases (as $a\rightarrow -\infty $ and $%
b\rightarrow \infty $). The left Fourier derivative agrees with the standard derivative for all
integer orders $\alpha $. However, if we use it for our problem, we cannot
find solutions (numerically) to Equation (\ref{steady1}), except for integer
orders $\alpha =2,4$ (and a negative sign is requred for $\alpha =4$). 

\begin{figure}[]
\begin{center}
\includegraphics[width=4.3cm]{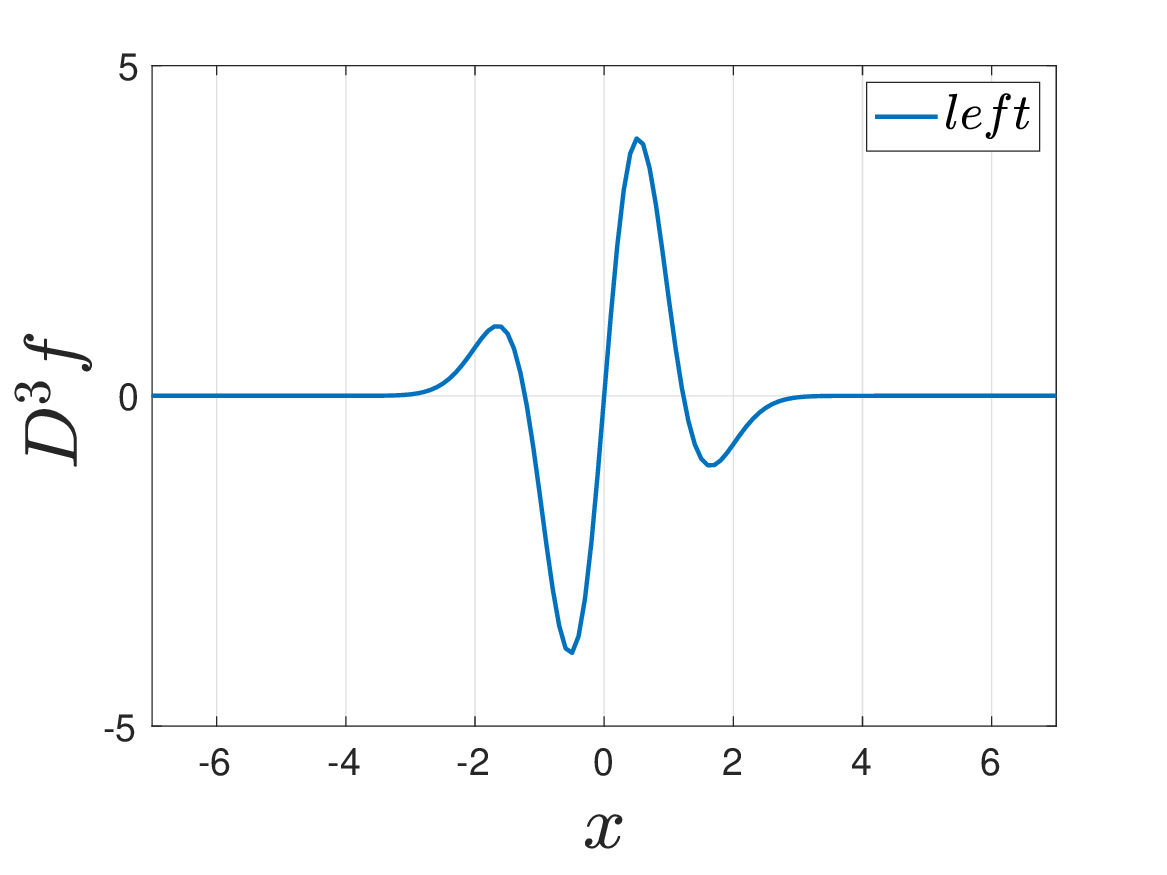} 
\includegraphics[width=4.3cm]{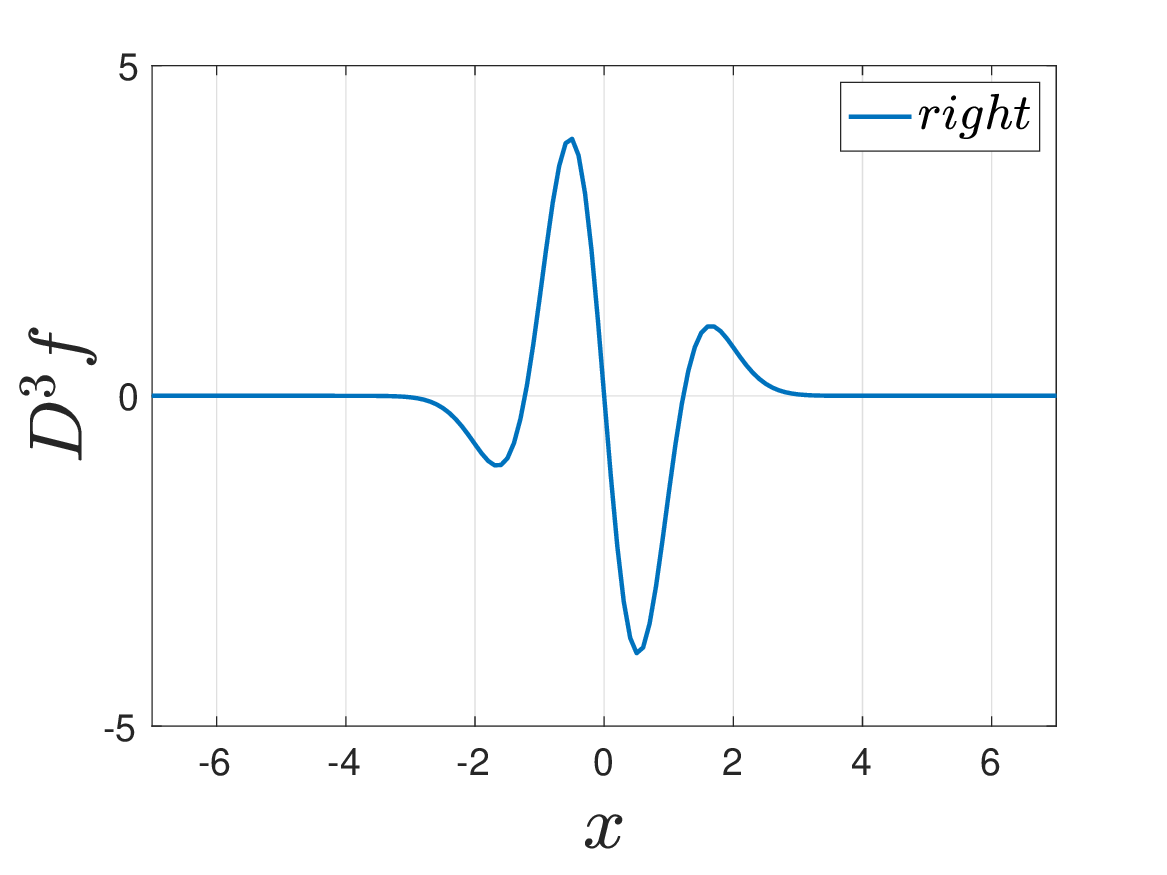} %
\includegraphics[width=4.3cm]{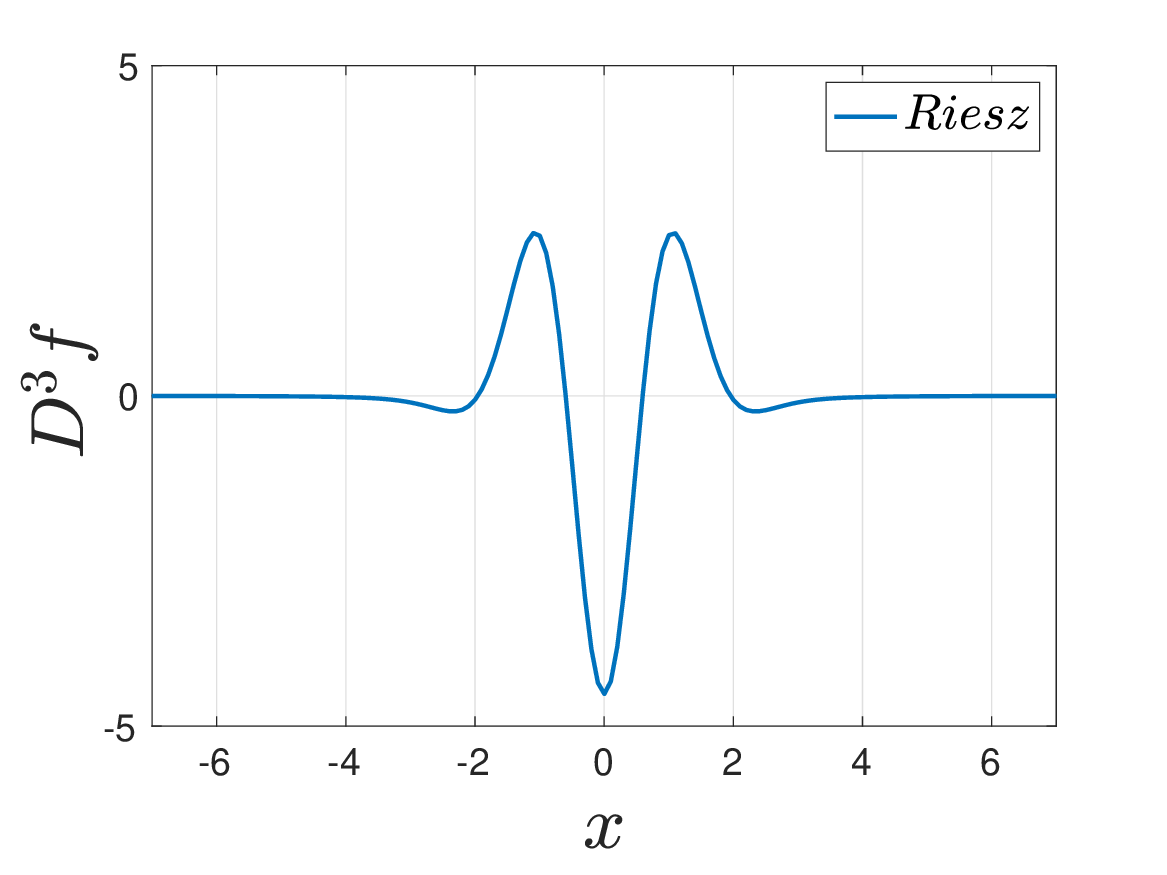} %
\includegraphics[width=4.3cm]{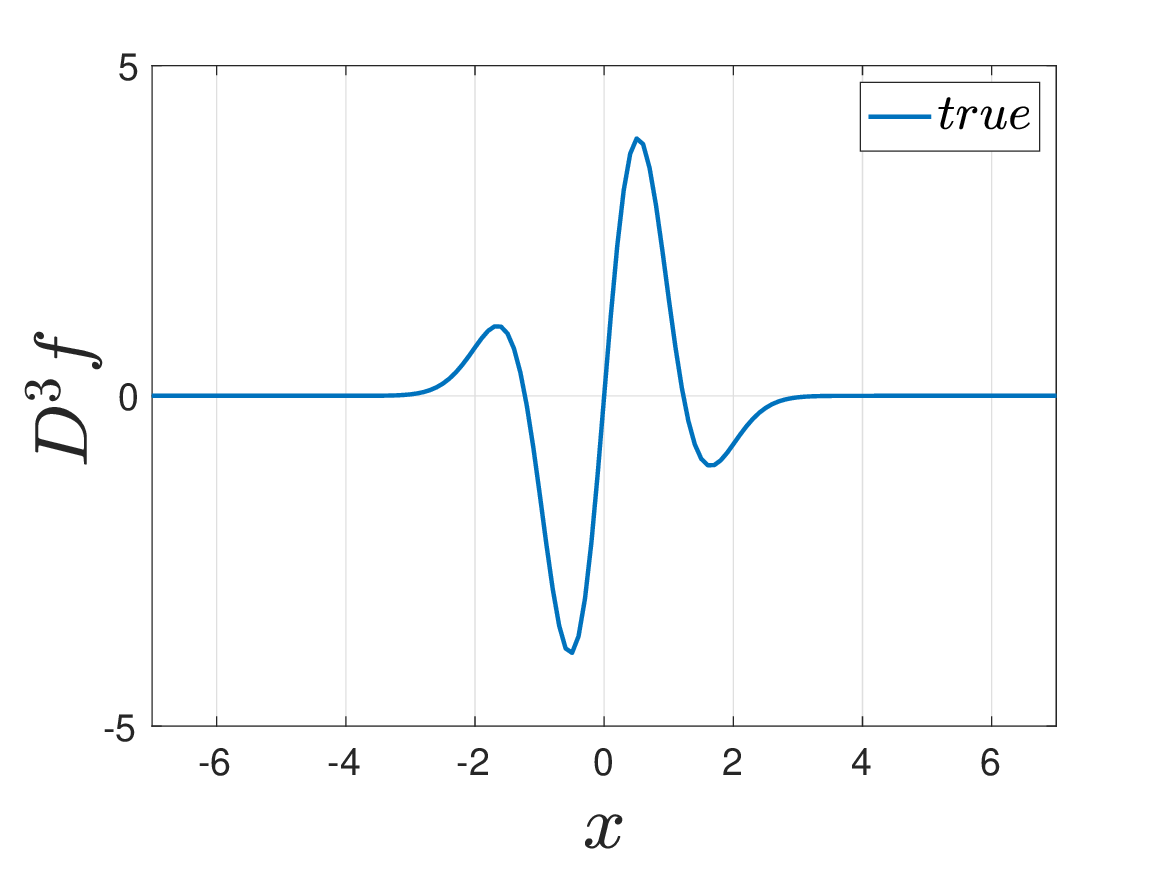}
\end{center}
\caption{$\protect\alpha=3$ derivatives of $f(x)=e^{-x^2}$: in sequence,
the left, the right, the Riesz and the
true 3rd derivative (which coincides with the left one).}
\label{alpha3}
\end{figure}

In order to get a visual feeling for the left, right, and Riesz (section \ref{sub:riesz}) derivatives, we demonstrate each applied to a Gaussian function, for $\alpha=3$, along with the standard third derivative, in Figure \ref{alpha3}. Here we see that the left and right derivatives are mirror left-right images and the Riesz derivative is a kind of ``symmetrized'' average of these two. Also note that the left derivative is the same as the standard derivative.

\bibliographystyle{apsrev4-1}
 \bibliography{Bibfile}

\end{document}